\documentstyle[10pt,epsf,epsfig,amssymb,amsfonts,amsmath,amsthm,cite,dp_delphititle,lineno]{dp_delphi}
\makeindex
\def\DpPaperGroup{EP}
\def\DpPaperRef{2002-088}
\def\DpDate{2 October 2002}
\def\DpAuthors{DELPHI Collaboration}
\def\DpSubmit{(Accepted by Eur. Phys. J. C)}
\def\DpTitle{{\boldmath
     b-tagging in DELPHI at LEP 
      }}
\def\DpComment{}
\def\DpEMail{ }

\def\ZP{Z.\ Phys.\ {\bf C}}
\def\PL{Phys.\ Lett.\ {\bf B}}

\def\be{\begin{equation}}
\def\ee{\end{equation}}
\def\bea{\begin{eqnarray}}
\def\eea{\end{eqnarray}}

\newcommand{\ba}{\begin{array}}
\newcommand{\ea}{\end{array}}
\newcommand{\bc}{\begin{center}}
\newcommand{\ec}{\end{center}}
\newcommand{\bt}{\begin{tabular}}
\newcommand{\et}{\end{tabular}}
\newcommand{\beq}{\begin{eqnarray}}
\newcommand{\eeq}{\end{eqnarray}}
\newcommand{\bes}{\begin{eqnarray*}}
\newcommand{\ees}{\end{eqnarray*}}

\begin{document}
\makeatletter
\makeatother
\begin{titlepage}
\pagenumbering{roman}
\CERNpreprint{\DpPaperGroup}{\DpPaperRef} 
\date{{\small\DpDate}} 
\title{\DpTitle} 
\address{\DpAuthors} 
\begin{shortabs} 
\noindent
\noindent
 The standard method used for tagging $b$-hadrons in 
 the DELPHI experiment at the CERN LEP Collider is discussed in detail. 
 The main ingredient of $b$-tagging is the impact parameters of tracks,
 which relies mostly on the vertex detector. Additional information,
 such as the mass of particles associated to a secondary vertex,
 significantly improves the selection efficiency and the background
 suppression. The paper describes various discriminating
 variables used for the tagging and the procedure of their combination.
 In addition, applications of $b$-tagging to some physics analyses, 
 which depend crucially on the performance and reliability 
 of $b$-tagging, are described briefly.
\end{shortabs}
\vfill
\begin{center}
\DpSubmit \ \\ 
\DpComment \ \\
\DpEMail \ \\
\end{center}
\vfill
\clearpage
\headsep 10.0pt
\addtolength{\textheight}{10mm}
\addtolength{\footskip}{-5mm}
\begingroup
%
\newcommand{\DpName}[2]{\hbox{#1$^{\ref{#2}}$},\hfill}
\newcommand{\DpNameTwo}[3]{\hbox{#1$^{\ref{#2},\ref{#3}}$},\hfill}
\newcommand{\DpNameThree}[4]{\hbox{#1$^{\ref{#2},\ref{#3},\ref{#4}}$},\hfill}
\newskip\Bigfill \Bigfill = 0pt plus 1000fill
\newcommand{\DpNameLast}[2]{\hbox{#1$^{\ref{#2}}$}\hspace{\Bigfill}}
%
\footnotesize
\noindent
\DpName{J.Abdallah}{LPNHE}
\DpName{P.Abreu}{LIP}
\DpName{W.Adam}{VIENNA}
\DpName{T.Adye}{RAL}
\DpName{P.Adzic}{DEMOKRITOS}
\DpName{T.Albrecht}{KARLSRUHE}
\DpName{T.Alderweireld}{AIM}
\DpName{R.Alemany-Fernandez}{CERN}
\DpName{T.Allmendinger}{KARLSRUHE}
\DpName{P.P.Allport}{LIVERPOOL}
% ...
\normalsize
\endgroup
\titlefoot{Department of Physics and Astronomy, Iowa State
     University, Ames IA 50011-3160, USA
    \label{AMES}}
\titlefoot{Physics Department, Universiteit Antwerpen,
     Universiteitsplein 1, B-2610 Antwerpen, Belgium \\
     \indent~~and IIHE, ULB-VUB,
     Pleinlaan 2, B-1050 Brussels, Belgium \\
     \indent~~and Facult\'e des Sciences,
     Univ. de l'Etat Mons, Av. Maistriau 19, B-7000 Mons, Belgium
    \label{AIM}}
\titlefoot{Physics Laboratory, University of Athens, Solonos Str.
     104, GR-10680 Athens, Greece
    \label{ATHENS}}
\titlefoot{Department of Physics, University of Bergen,
     All\'egaten 55, NO-5007 Bergen, Norway
    \label{BERGEN}}
\titlefoot{Dipartimento di Fisica, Universit\`a di Bologna and INFN,
     Via Irnerio 46, IT-40126 Bologna, Italy
    \label{BOLOGNA}}
\titlefoot{Centro Brasileiro de Pesquisas F\'{\i}sicas, rua Xavier Sigaud 150,
     BR-22290 Rio de Janeiro, Brazil \\
     \indent~~and Depto. de F\'{\i}sica, Pont. Univ. Cat\'olica,
     C.P. 38071 BR-22453 Rio de Janeiro, Brazil \\
     \indent~~and Inst. de F\'{\i}sica, Univ. Estadual do Rio de Janeiro,
     rua S\~{a}o Francisco Xavier 524, Rio de Janeiro, Brazil
    \label{BRASIL}}
\titlefoot{Coll\`ege de France, Lab. de Physique Corpusculaire, IN2P3-CNRS,
     FR-75231 Paris Cedex 05, France
    \label{CDF}}
\titlefoot{CERN, CH-1211 Geneva 23, Switzerland
    \label{CERN}}
\titlefoot{Institut de Recherches Subatomiques, IN2P3 - CNRS/ULP - BP20,
     FR-67037 Strasbourg Cedex, France
    \label{CRN}}
\titlefoot{Now at Universita dell'Insubria in Como, Dip.to di Scienze CC.FF.MM`
     via Vallegio 11, 1-22100 Como, Italy
    \label{COMO}}
\titlefoot{Now at DESY-Zeuthen, Platanenallee 6, D-15735 Zeuthen, Germany
    \label{DESY}}
\titlefoot{Institute of Nuclear Physics, N.C.S.R. Demokritos,
     P.O. Box 60228, GR-15310 Athens, Greece
    \label{DEMOKRITOS}}
\titlefoot{Now at Fermilab (FNAL), Kirk and Pine Streets, P.O. Box 500,
     Batavia, IL 60510     
    \label{FERMILAB}}
\titlefoot{FZU, Inst. of Phys. of the C.A.S. High Energy Physics Division,
     Na Slovance 2, CZ-180 40, Praha 8, Czech Republic
    \label{FZU}}
\titlefoot{Dipartimento di Fisica, Universit\`a di Genova and INFN,
     Via Dodecaneso 33, IT-16146 Genova, Italy
    \label{GENOVA}}
\titlefoot{Institut des Sciences Nucl\'eaires, IN2P3-CNRS, Universit\'e
     de Grenoble 1, FR-38026 Grenoble Cedex, France
    \label{GRENOBLE}}
\titlefoot{Helsinki Institute of Physics, HIP,
     P.O. Box 9, FI-00014 Helsinki, Finland
    \label{HELSINKI}}
\titlefoot{Joint Institute for Nuclear Research, Dubna, Head Post
     Office, P.O. Box 79, RU-101 000 Moscow, Russian Federation
    \label{JINR}}
\titlefoot{Institut f\"ur Experimentelle Kernphysik,
     Universit\"at Karlsruhe, Postfach 6980, DE-76128 Karlsruhe,
     Germany
    \label{KARLSRUHE}}
\titlefoot{Institute of Nuclear Physics,Ul. Kawiory 26a,
     PL-30055 Krakow, Poland
    \label{KRAKOW1}}
\titlefoot{Faculty of Physics and Nuclear Techniques, University of Mining
     and Metallurgy, PL-30055 Krakow, Poland
    \label{KRAKOW2}}
\titlefoot{Universit\'e de Paris-Sud, Lab. de l'Acc\'el\'erateur
     Lin\'eaire, IN2P3-CNRS, B\^{a}t. 200, FR-91405 Orsay Cedex, France
    \label{LAL}}
\titlefoot{School of Physics and Chemistry, University of Landcaster,
     Lancaster LA1 4YB, UK
    \label{LANCASTER}}
\titlefoot{LIP, IST, FCUL - Av. Elias Garcia, 14-$1^{o}$,
     PT-1000 Lisboa Codex, Portugal
    \label{LIP}}
\titlefoot{Department of Physics, University of Liverpool, P.O.
     Box 147, Liverpool L69 3BX, UK
    \label{LIVERPOOL}}
\titlefoot{LPNHE, IN2P3-CNRS, Univ.~Paris VI et VII, Tour 33 (RdC),
     4 place Jussieu, FR-75252 Paris Cedex 05, France
    \label{LPNHE}}
\titlefoot{Department of Physics, University of Lund,
     S\"olvegatan 14, SE-223 63 Lund, Sweden
    \label{LUND}}
\titlefoot{Universit\'e Claude Bernard de Lyon, IPNL, IN2P3-CNRS,
     FR-69622 Villeurbanne Cedex, France
    \label{LYON}}
\titlefoot{Dipartimento di Fisica, Universit\`a di Milano and INFN-MILANO,
     Via Celoria 16, IT-20133 Milan, Italy
    \label{MILANO}}
\titlefoot{Dipartimento di Fisica, Univ. di Milano-Bicocca and
     INFN-MILANO, Piazza della Scienza 2, IT-20126 Milan, Italy
    \label{MILANO2}}
\titlefoot{IPNP of MFF, Charles Univ., Areal MFF,
     V Holesovickach 2, CZ-180 00, Praha 8, Czech Republic
    \label{NC}}
\titlefoot{NIKHEF, Postbus 41882, NL-1009 DB
     Amsterdam, The Netherlands
    \label{NIKHEF}}
\titlefoot{National Technical University, Physics Department,
     Zografou Campus, GR-15773 Athens, Greece
    \label{NTU-ATHENS}}
\titlefoot{Physics Department, University of Oslo, Blindern,
     NO-0316 Oslo, Norway
    \label{OSLO}}
\titlefoot{Dpto. Fisica, Univ. Oviedo, Avda. Calvo Sotelo
     s/n, ES-33007 Oviedo, Spain
    \label{OVIEDO}}
\titlefoot{Department of Physics, University of Oxford,
     Keble Road, Oxford OX1 3RH, UK
    \label{OXFORD}}
\titlefoot{Dipartimento di Fisica, Universit\`a di Padova and
     INFN, Via Marzolo 8, IT-35131 Padua, Italy
    \label{PADOVA}}
\titlefoot{Rutherford Appleton Laboratory, Chilton, Didcot
     OX11 OQX, UK
    \label{RAL}}
\titlefoot{Dipartimento di Fisica, Universit\`a di Roma II and
     INFN, Tor Vergata, IT-00173 Rome, Italy
    \label{ROMA2}}
\titlefoot{Dipartimento di Fisica, Universit\`a di Roma III and
     INFN, Via della Vasca Navale 84, IT-00146 Rome, Italy
    \label{ROMA3}}
\titlefoot{Now at Inst. of Physical and Chemical Research RIKEN,
     2-1 Hirosawa, Wako-shi, Saitama 351-0198, Japan
    \label{RIKEN}} 
\titlefoot{DAPNIA/Service de Physique des Particules,
     CEA-Saclay, FR-91191 Gif-sur-Yvette Cedex, France
    \label{SACLAY}}
\titlefoot{Instituto de Fisica de Cantabria (CSIC-UC), Avda.
     los Castros s/n, ES-39006 Santander, Spain
    \label{SANTANDER}}
\titlefoot{Inst. for High Energy Physics, Serpukov
     P.O. Box 35, Protvino, (Moscow Region), Russian Federation
    \label{SERPUKHOV}}
\titlefoot{J. Stefan Institute, Jamova 39, SI-1000 Ljubljana, Slovenia
     and Laboratory for Astroparticle Physics,\\
     \indent~~Nova Gorica Polytechnic, Kostanjeviska 16a, SI-5000 Nova Gorica, Slovenia, \\
     \indent~~and Department of Physics, University of Ljubljana,
     SI-1000 Ljubljana, Slovenia
    \label{SLOVENIJA}}
\titlefoot{Fysikum, Stockholm University,
     Box 6730, SE-113 85 Stockholm, Sweden
    \label{STOCKHOLM}}
\titlefoot{Dipartimento di Fisica Sperimentale, Universit\`a di
     Torino and INFN, Via P. Giuria 1, IT-10125 Turin, Italy
    \label{TORINO}}
\titlefoot{INFN,Sezione di Torino, and Dipartimento di Fisica Teorica,
     Universit\`a di Torino, Via P. Giuria 1,\\
     \indent~~IT-10125 Turin, Italy
    \label{TORINOTH}}
\titlefoot{Now at Institute of Particule Physics of Canada, Univ. of Toronto,
     Toronto, Ontario, Canada M5S1A7
    \label{TORONTO}} 
\titlefoot{Dipartimento di Fisica, Universit\`a di Trieste and
     INFN, Via A. Valerio 2, IT-34127 Trieste, Italy \\
     \indent~~and Istituto di Fisica, Universit\`a di Udine,
     IT-33100 Udine, Italy
    \label{TU}}
\titlefoot{Univ. Federal do Rio de Janeiro, C.P. 68528
     Cidade Univ., Ilha do Fund\~ao
     BR-21945-970 Rio de Janeiro, Brazil
    \label{UFRJ}}
\titlefoot{Department of Radiation Sciences, University of
     Uppsala, P.O. Box 535, SE-751 21 Uppsala, Sweden
    \label{UPPSALA}}
\titlefoot{IFIC, Valencia-CSIC, and D.F.A.M.N., U. de Valencia,
     Avda. Dr. Moliner 50, ES-46100 Burjassot (Valencia), Spain
    \label{VALENCIA}}
\titlefoot{Institut f\"ur Hochenergiephysik, \"Osterr. Akad.
     d. Wissensch., Nikolsdorfergasse 18, AT-1050 Vienna, Austria
    \label{VIENNA}}
\titlefoot{Inst. Nuclear Studies and University of Warsaw, Ul.
     Hoza 69, PL-00681 Warsaw, Poland
    \label{WARSZAWA}}
\titlefoot{Fachbereich Physik, University of Wuppertal, Postfach
     100 127, DE-42097 Wuppertal, Germany \\
\noindent
{$^\dagger$~deceased}
    \label{WUPPERTAL}}
\addtolength{\textheight}{-10mm}
\addtolength{\footskip}{5mm}
\clearpage
\headsep 30.0pt
\end{titlepage}
%
\pagenumbering{arabic} 
\setcounter{footnote}{0} %
\large
\section{Introduction}

The study of heavy $b$- and $c$- quarks is one of the most interesting
subjects in experimental High Energy Physics, directly related to the
verification of the Standard Model (SM) and the search for its possible
violations. Where these may occur is very
model-dependent, but it may well be that the third generation particles will
provide some important clues to new effects. This is a large part of the
motivation for studying $b$-quarks at LEP, where top-quark pair production is
kinematically inaccessible. It is thus important to have
algorithms for selecting events with $b$-quarks while keeping backgrounds 
small.
Efficiency and purity or background rejection are important parameters of these
techniques.
Because searches for such deviations from the SM often involve 
precision measurements, it is crucial to have a very well understood 
and monitored $b$-tagging algorithm.

A further reason for selecting $b$-quarks is the search for the Higgs boson. 
For the SM Higgs with  mass of relevance for LEP and the Tevatron, the
predominant decay mode is to $b\bar{b}$ pairs. Thus tagging 
$b$-jets
provides a valuable means of selecting candidates while reducing backgrounds
to low levels, thereby enabling searches to achieve high sensitivity.

In this paper the $b$-tagging technique developed 
for the DELPHI
experiment at the LEP electron-positron collider is described.
LEP ran at centre-of-mass
energies around the $Z$ (91 GeV) over the period 1989 to 1995, and then at
higher energies up to 208 GeV, before being turned off in 2000. 
Much of the technique used here would be applicable, with
suitable modifications, in other experimental situations.

The lifetimes of $b$-hadrons are around 1.6 ps. This means that flight 
distances 
are of order 3 mm for a 35 GeV $b$-hadron,  this being a typical energy
in a 2-jet event at the $Z$, or in a 4-jet event at LEP2 energies.
Correspondingly the decay tracks from a $b$-hadron have non-zero 
impact parameters\footnote{See Section \ref{sec11} for more detailed
definitions and discussion of impact parameters.},
i.e. when extrapolated backward in space they do not pass exactly through the
beam interaction region. The scale of these impact parameters is $c\tau
\approx$ 400
$\mu$m. This is to be compared with the DELPHI experimental resolution
$\sigma$ of about 
\begin{equation}
\sigma = 27 \oplus 63/(p\, \sin^{3/2}\theta)\, \mu \mbox{m}
\label{firsteqn}
\end{equation}
where $p$  and $\theta$ are the
momentum (in GeV/c) and the polar angle of the track. 
The symbol $\oplus$ denotes the quadratic sum of terms. Eqn.~(\ref{firsteqn}) 
is for the impact parameter (IP) in the plane perpendicular to 
the beam; along the beam
direction, the resolution is slightly worse. 
Because the micro-vertex detector is crucial for achieving this accuracy 
in IP  measurements, it is described in Section \ref{sect:vd}.

The impact parameters provide the main variable for $b$-tagging.
For all the tracks in a jet, the observed impact parameters and resolutions are
combined into a single variable, the lifetime probability,
which measures the consistency with the hypothesis that all tracks come directly
from the primary vertex. For events without long-lived particles,       
this variable should be uniformly distributed between zero and unity. In
contrast, for $b$-jets it has predominantly small values.
Details of how this variable is
constructed are elaborated in Section \ref{sec2}.

Other features of the event are also sensitive to $b$-quarks, and some of them
are also used together with the IP information to construct a
`combined tag'. For example, $b$-hadrons have a 10\% probability of decaying to
electrons or muons, and these often have a
transverse momentum with respect to the $b$-jet axis of around 1 GeV/c or
larger. On its
own, the high-$p_T$ lepton tag would have too low an efficiency for many 
$b$-quark
studies, but the presence of such a lepton is useful information to combine
with the IP measurements. The combined tag also makes use of other variables
which have significantly different distributions for $b$-quark and for other
events, e.g. the charged particle rapidities with respect to the jet axis. 
Further details
on these variables and the way in which they are combined are given in
Section \ref{sec3}. The combined tag including the lifetime probability and
secondary vertex mass, rapidities and fractional energy (described in Section
\ref{sec23})
was used for the measurements at the $Z$ (see Sections \ref{sect:Rb} --
\ref{sect:mb}).
For most LEP2 $b$-tagging analyses, the transverse momentum missing at the 
secondary vertex and the transverse momentum with respect to the jet axis  of 
any electron or muon were also used in the combination.

The combination method used is optimal for uncorrelated variables. The extent
to which it is possible to improve on the `combined tag', for example by using
extra information such as the jet energy, is investigated in Section 
\ref{sect:equalise}. 
The resulting `equalised
tag' was used in the Higgs search at LEP2 (see Section \ref{Higgs}).

Section \ref{mass} contains some technical aspects of the 
$b$-tagging. In particular it describes 
some modifications that were required to the
physics generators of the Monte Carlo simulation.

Finally, some physics studies for which $b$-tagging plays a crucial role are
outlined in Section \ref{applications}. First there is the measurement of the 
fraction of hadronic $Z$ decays which contain $b$-quarks (see Section 
\ref{sect:Rb}). A precise
measurement of this quantity requires high efficiency tagging, while keeping
down the backgrounds from other quarks in order to reduce the systematic
errors. 
This is followed by applications of $b$-tagging to the measurement 
of the production rate of events with 4 $b$-jets, and of the $b$-hadron 
charged decay multiplicity. 
Section \ref{sect:mb} describes a measurement of the
$b$-fraction in 3-jet events, which is sensitive to the mass of the $b$-quark.
This uses anti $b$-tagging to select light quark events.
Finally the crucial reliance on $b$-tagging of the
search for the Higgs is described in Section \ref{Higgs}. 

\section{The DELPHI Vertex Detector}

\label{sect:vd}

\subsection{Overview}

The silicon vertex detectors of the DELPHI experiment have
undergone various upgrades throughout the lifetime of the experiment.
For the statistics collected in 1991-1993 it provided measurements
in the transverse ($R\phi$) plane\footnote{DELPHI uses
a cylindrical polar co-ordinate system, with the $z$ axis along the beam
direction (and the magnetic field axis).  $R$ and $\phi$ are the radial
and azimuthal co-ordinates in the transverse plane, $\theta$ is the polar
angle with respect to the beam axis. The Cartesian co-ordinates $x$
and $y$ are horizontal and vertical respectively.} only \cite{nim1}.

\begin{figure}[htb]
\epsfig{file=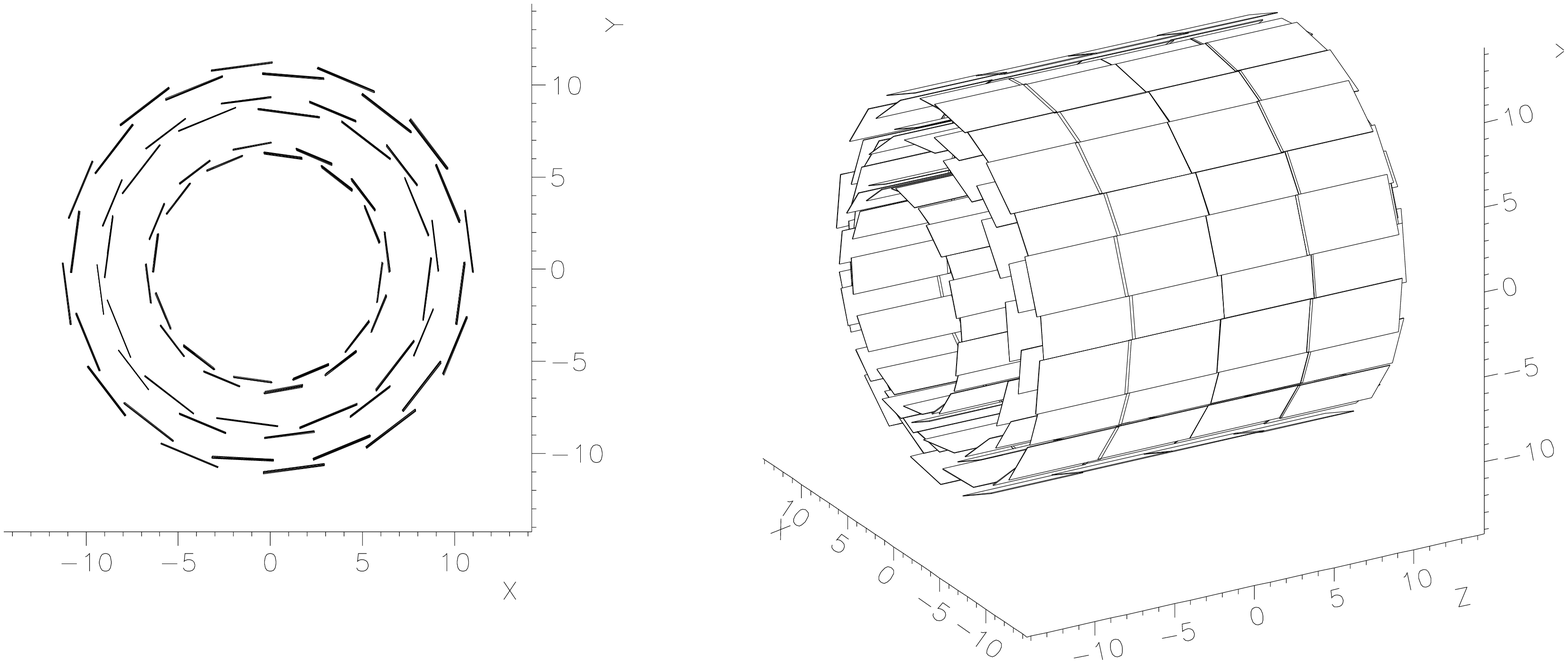,width=1.0\textwidth}
\caption{Schematic cross sections of the Double Sided Vertex Detector
 in {\bf (a)} the transverse 
($R\phi$) view and {\bf (b)}~a three-dimensional view. Dimensions are in cm.}
\label{dsvd}
\end{figure}

The DELPHI Double Sided Vertex Detector (DSVD)~\cite{nimdsvd} 
was installed in the experiment in
early 1994 and by the end of the $Z$ running at LEP had contributed
to the reconstruction and analysis of approximately 2 million
$Z$ decays.  
Two of its three layers were equipped with double-sided orthogonal readouts,
thereby upgrading the IP and vertexing capabilities
by adding information from the longitudinal ($Rz$) plane.
The extra coordinate helps to associate tracks to vertices where the
single $R\phi$ view alone might have ambiguities.  This upgrade led
to about a $30 \%$ improvement in the $b$-tagging efficiency at 
fixed purity. The geometrical layout of the DSVD is shown in figure~\ref{dsvd}.
The three layers, termed Closer, Inner, and Outer, were at average
radii of 6.3, 9.0 and 10.9 cm respectively, with the Outer and Closer layers
instrumented with the double-sided orthogonal readout.  
The three-layer polar angular 
coverage was between $44^{\circ}$ and $136^{\circ}$, with the Closer Layer
providing additional coverage in anticipation of the subsequent SiT upgrade 
described below.  The transverse view displays the large degree of overlap
(up to $20\%$ of the sensitive region in the Inner Layer), which was
an important ingredient for the alignment.
The average thickness of each silicon module was
$0.5 \%$ of a radiation length. The $z$ readout was routed via an integrated 
double metal layer, thus adding negligible  extra material in the barrel region,
and helping to keep multiple scattering to a minimum.

\begin{figure}[htb]
\epsfig{file=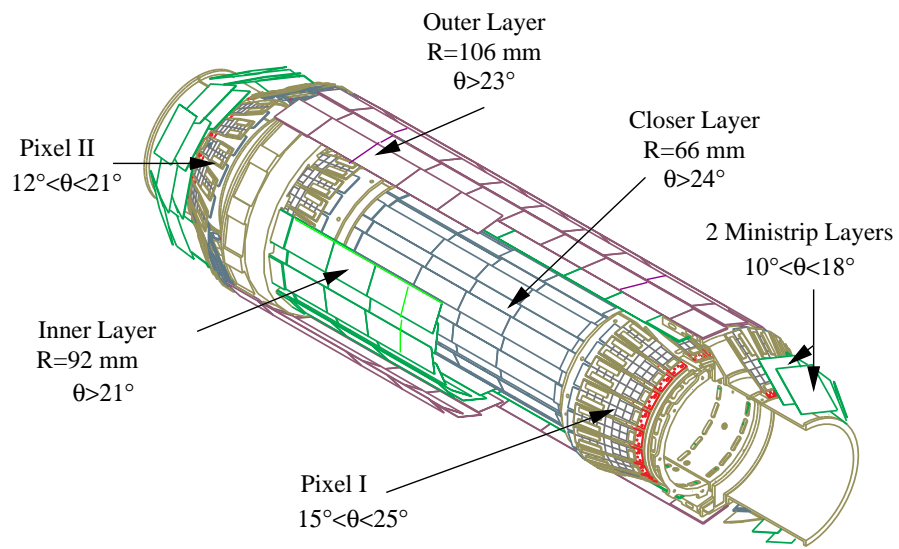,width=1.0\textwidth}
\caption{Schematic view of the Silicon Tracker}
\label{SiT}
\end{figure}

\begin{figure}[htb]
\epsfig{file=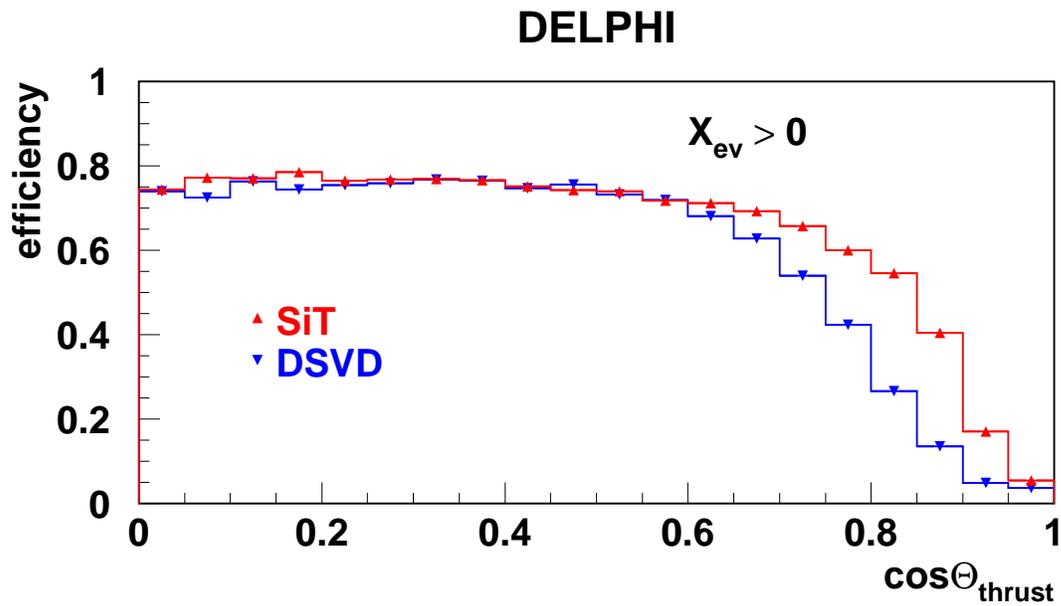,width=1.0\textwidth}
\caption{Selection efficiency of $Z \to b \bar{b}$ events 
versus $\cos \Theta_{thrust}$ using the combined event $b$-tagging 
variable $X_{ev} > 0$ for the Silicon Tracker and the Double Sided Vertex
Detector. ($X_{ev}$ is defined in Section \ref{descrip}.)
The extra coverage provided by the Silicon Tracker in the forward 
direction is clearly visible.}
\label{chiarathetaplot}
\end{figure}

The DELPHI Silicon Tracker (SiT)~\cite{nimsit} was a further upgrade for the
physics requirements at LEP2, and the barrel part
relevant for $b$-tagging was fully installed in 1996.
Physics objectives of LEP2, such as the measurement
of four-fermion processes and the searches for the Higgs
boson or for super-symmetric particles,
required a larger polar angle coverage than at LEP1.
The design goal was to achieve an equivalent $b$-tagging performance
to the DELPHI DSVD, and in addition to extend this to around
$25^{\circ}$ in $\theta$, after which the $b$-tagging capabilities were
limited by multiple scattering in the beam-pipe.  The
SiT also incorporated end-caps of mini-strip and
pixel detectors for tracking in the forward region~\cite{nimsit,vft}.
The geometrical layout of the SiT is shown in figure~\ref{SiT}.
The radii of the layers were similar to the DSVD, but the Outer
and Inner layers were doubled in length to provide the
extra angular coverage.  The Closer Layer was double-sided,
the Inner Layer was double-sided for $21^{\circ} < \theta < 44^{\circ}$ (and
the corresponding backward region)
and single-sided in the centre, and the Outer Layer provided
$R\phi$ and $Rz$ measurements from its crossed detector 
arrangement~\cite{nimsit}.
The impact of this detector on $b$-tagging is shown in
figure~\ref{chiarathetaplot}.

\subsection{Alignment and Performance}
\label{align}

$b$-tagging quality relies on excellent alignment of the vertex detector.
The starting point of the alignment
was the information from an optical and mechanical survey before
installation.  This was refined with the information from tracks from
$Z$ decays, using a stand-alone
procedure where the momentum of the track was the only information
taken from the rest of the DELPHI detector.  The precision of the vertex 
detector hits has allowed a number of important effects to be identified,
including some  common to all LEP vertex detectors and certain
previously unmeasured properties of silicon detectors.  They
include:

\begin{itemize}
\item{coherent deformations, such as a torsion or shear of the
entire structure;}
\item{bowing of the silicon modules due to the different response
of the silicon and the module support to changes in temperature
and humidity;}
\item{barycentric shift effects, whereby the centres of gravity
of the charge clouds of electrons and holes in the silicon
do not correspond to the mid-plane of the detector, nor to each other;}
\item{acollinearity of the LEP electron and positron beams
leading to lepton pairs from $Z$ decays which cannot be assumed to
be back-to-back in the alignment procedure.}
\end{itemize}
More details can be found in \cite{alignnote}. The precise vertex 
detector alignment has also led to better understanding
of other detectors, such as the TPC, the track distortions of which were
corrected. 

\begin{figure}[htb]
\epsfig{file=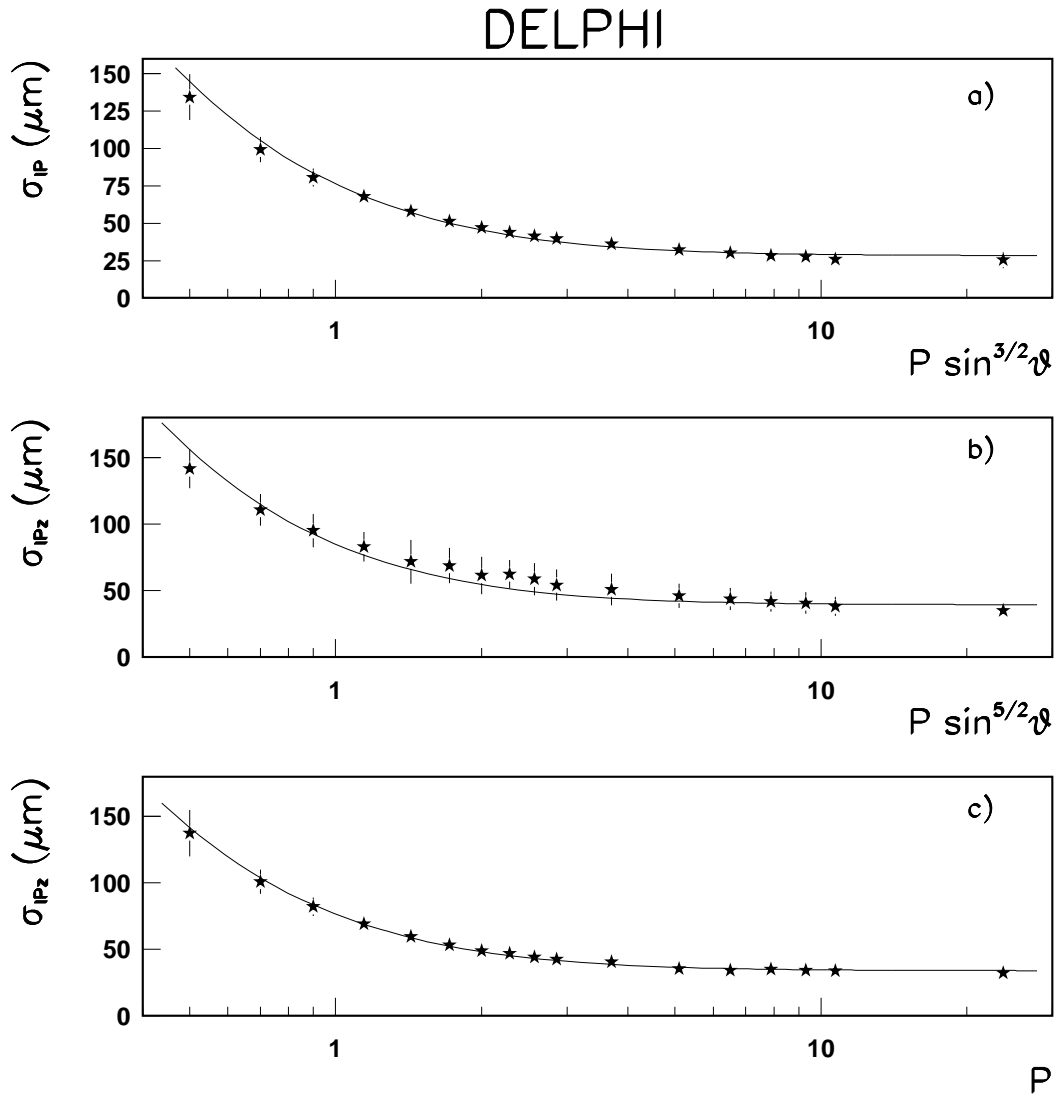,width=1.0\textwidth}
\caption{The $R\phi$ IP uncertainty as a function of 
${\rm p~sin^{\frac{3}{2}}\theta}$ (upper plot), the $Rz$ IP uncertainty
as a function of ${\rm p~sin^{\frac{5}{2}}\theta}$ (middle plot)
and the $Rz$ IP uncertainty as a function of $p$ for tracks
with $\theta = [ 80^{\circ}:100^{\circ} ] $, i.e. perpendicular 
to the beam direction (lower plot).
The data are from the Double Sided Vertex Detector at the $Z$.
The curves are parameterisations of a constant intrinsic resolution term
and a momentum-dependent multiple scattering contribution. Momenta
are in Gev/c.}
\label{ipres}
\end{figure}

The ultimate performance of the vertex detector with respect to $b$-tagging
can be measured by the IP resolution, which in the $R\phi$ plane 
can be parametrised by:


\begin{eqnarray}
\label{resrp1}
\sigma_{R\phi} &=& 27 \oplus \, 
\frac{63}{{\rm p~sin^{\frac{3}{2}}\theta}}\ \mu{\rm m}
\label{resrz1}
\end{eqnarray}
with $p$ in GeV/c. The IP resolution in the $Rz$ plane for 
two typical $\theta$ regions can be parametrised by:

\begin{eqnarray}
\sigma_{Rz} &=& 39 \oplus \, \frac{71}{\rm p}
\ \mu{\rm m}~({\rm for}~80^{\circ} < \theta < 90^{\circ})\\
\label{resrz2}
\sigma_{Rz} &=& 96 \oplus \, \frac{151}{\rm p} 
\ \mu{\rm m}~({\rm for}~45^{\circ} < \theta < 55^{\circ}),
\end{eqnarray}
These equations are the quadratic sums of a constant 
and a momentum dependent term,
corresponding to the intrinsic resolution and to the
multiple scattering contributions  respectively. For tracks coming from 
$b$-decays, these contributions are of similar magnitude.
Typical distributions of the IP resolutions as functions of 
momenta are shown in figure~\ref{ipres}.


\section{Lifetime Tagging}
\label{sec2}

$b$-hadrons in many aspects are significantly different from all other 
particles. They have a long lifetime, large mass, high decay multiplicity,
substantial leptonic branching rate, etc. 
The most important property
for the selection of $b$-hadrons is their lifetime.
Among the main features of lifetime tagging are a simple and transparent 
definition and ease of control, since it relies on 
a single measured quantity, the track IP.

In this section the definition of the main elements entering in the
lifetime tagging together with the principles of its construction are given.
This tagging
itself provides efficient separation of the $b$-quark from other
flavours, which is further enhanced by including additional variables (see
Section \ref{sec3}). 
The  method of lifetime tagging used by DELPHI was originally proposed
by the ALEPH collaboration\cite{aleph}.

\subsection{Impact Parameter}
\label{sec11}

The general 3-dimensional IP is the minimal distance between the 
estimated primary interaction point and the track trajectory. The 
decay of a long-lived particle
produces tracks with large impact parameters, which is not the case
for particles from the primary interaction. Lifetime tagging is based
on this difference.

\begin{figure}[htb]
      \epsfig{figure=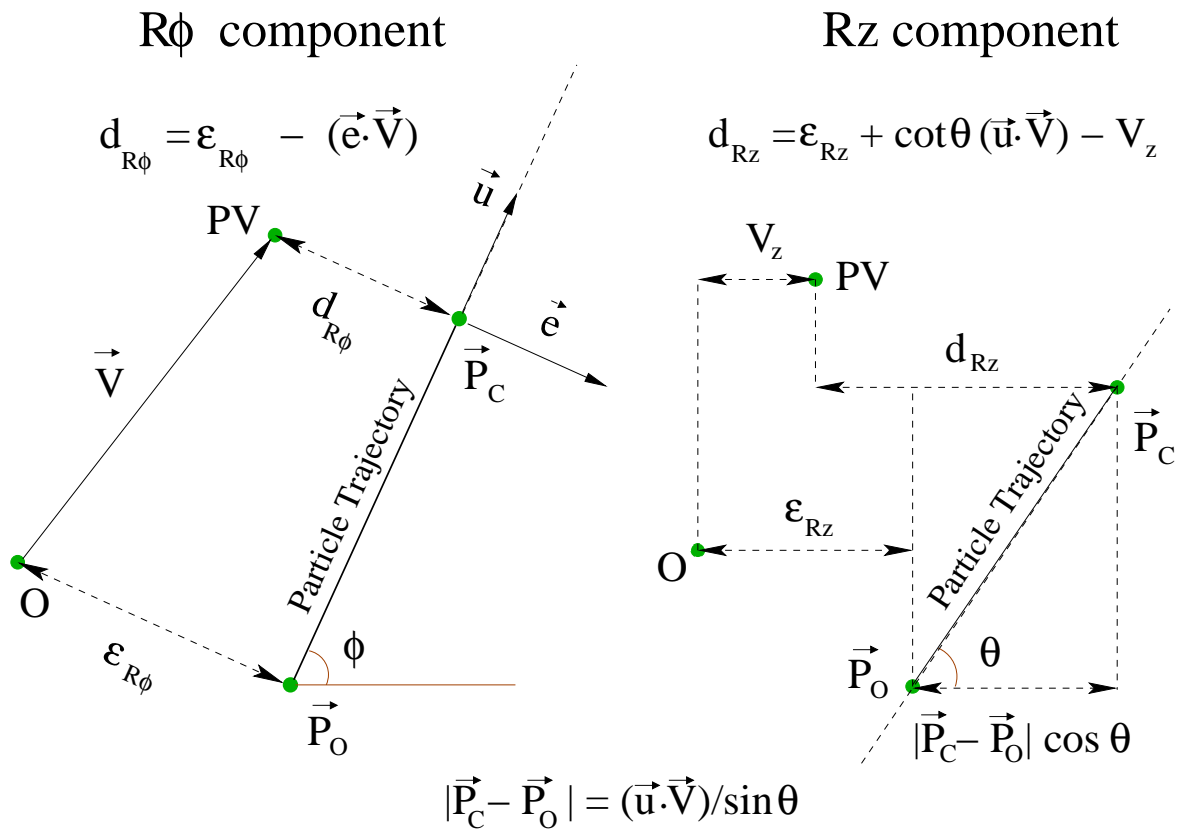,width=16cm}
    \caption[]{Definition of $R\phi$ and $Rz$ IP components.
\mbox{$\overrightarrow{u}$} is a unit vector along the track direction, and 
\mbox{$\overrightarrow{e}$} is another unit vector
in the $R\phi$ plane, perpendicular to \mbox{$\overrightarrow{u}$}. 
\mbox{$\overrightarrow{V}$} is a vector from the origin $O$
to the primary vertex PV. $P_0$ and $P_C$ are the points of closest approach 
in the $R\phi$
plane of the track trajectory to $O$ and to PV respectively. 
The diagrams show the projections onto the $R\phi$ and $Rz$ planes.
The IP components are $d_{R\phi}$ and $d_{Rz}$, while
$\varepsilon_{R\phi}$ and $\varepsilon_{Rz}$ are the corresponding components 
from $P_0$ to the origin. }
    \label{fig1}
\end{figure}

For $b$-tagging in DELPHI, a slightly different approach is adopted, with a
separation of the 3-dimensional information into $R\phi$ and $Rz$ components. 
The IP component in the $R\phi$ plane is defined as the minimal
distance between the primary vertex (PV) and the track trajectory  
projected onto the plane perpendicular to the beam direction. 
The point of the closest approach ($P_C$) of the track trajectory to the 
primary vertex in the $R\phi$ plane 
is also used to define the $Rz$ component of the IP. 
This is the difference
between the $z$-coordinates of the primary vertex and of the point $P_C$ 
(see fig.~\ref{fig1}).

According to these definitions, there are two ingredients in the IP 
computation: the parameters of the track trajectory, provided by the track fit, 
and the position of the primary interaction. The parameters of the track
trajectory are the track direction given by its polar and azimuthal angles 
($\theta, \phi$) at the point $P_0$ of the closest approach to the origin $O$;
 and ($\varepsilon_{R\phi}$, $\varepsilon_{Rz}$),
the equivalent of the IP components  but defined with respect to the origin 
$O$,
rather than with respect to the primary vertex. 
The reconstruction of the primary vertex is explained in the next section. 
In the approximation that the tracks can be
regarded  as straight lines between $P_0$ and $P_C$, the IP components
$d_{R\phi}$ and $d_{Rz}$ with respect to the primary vertex 
position $\overrightarrow{V}$ are calculated as:


\begin{eqnarray}
  \label{eq1}
  d_{R\phi} & = &  \varepsilon_{R\phi} - 
    (\overrightarrow{e} \cdot \overrightarrow{V}) \\
  \label{eq2}
  d_{Rz} & = &  \varepsilon_{Rz} + 
    \cot \theta (\overrightarrow{u} \cdot \overrightarrow{V})-V_z \nonumber \\
         & = & 
    \varepsilon_{Rz} - (\overrightarrow{l} \cdot \overrightarrow{V})
\end{eqnarray}
Here 
$\overrightarrow{u}$ is the unit vector along the track direction in the $R\phi$
plane: $\overrightarrow{u} = \{\cos \phi, \sin\phi,0\}$;
$\overrightarrow{e}$ is the unit vector perpendicular to the track
direction in $R\phi$ plane: $\overrightarrow{e} = \{\sin \phi, -\cos \phi,0\}$;
 and
$\overrightarrow{l} = \{-\cot \theta \cos \phi, -\cot \theta \sin\phi,1\}$.
Figure~\ref{fig1} illustrates these definitions of the IP components.

The main reason for the separation of the 3-dimensional IP into $R\phi$
and $Rz$ components is that the measurement of the particle 
trajectory in DELPHI is performed independently in these two planes with 
somewhat different precision (see eqns. (\ref{resrp1}) -- (\ref{resrz2})). 
Also the beam-spot is smaller in the transverse directions.
In addition, there are 3 sensitive layers 
of vertex detector in the $R\phi$ plane and only 2 layers in the $Rz$ plane;
the fraction of tracks with wrong hit association in the $Rz$ plane 
is thus higher. The separate treatment of the IP components provides
the freedom to reject bad measurements in the $Rz$ plane,
while keeping  useful $R\phi$ information. Finally, the data before 1994
were taken with the 2-dimensional vertex detector
providing track measurements in the $R\phi$ plane only. The separate use of 
the $R\phi$ and $Rz$ information is one
of the crucial points of our tagging algorithm, significantly influencing
its structure. 

\subsection{Primary Vertex}
\label{sec12}

The primary vertex is reconstructed for each event using a set of selected
tracks  and the beam-spot position. The beam-spot is the zone of 
intersection of the two colliding beams of LEP. It has a small size in the 
$R\phi$ plane ($\sigma_x \simeq 150 \mu$m, $\sigma_y$ less than $10\mu$m), 
while it is several millimetres long along the beam direction. 
It is relatively 
stable within a fill, and so can be used as a constraint for the 
primary vertex fit. 

The beam-spot is measured using events which have
a vertex formed by at least 3 tracks with hits in the silicon strip
detectors. These vertices are used to fit the position in 3 dimensions and
also the $x$ and $z$ size 
of the interaction region in time periods of around 20 minutes. 
The size of the interaction region in $y$ is not fitted, 
because it is smaller than the corresponding position error,
and the value $\sigma_y = 10$ $\mu$m is used.

The PV position is obtained by minimising the $\chi^2$ function:
\begin{equation}
\chi^2(\overrightarrow{V}) = \sum_{a} \sum_{\alpha,\beta=1,2}
d^a_{\alpha} (S^{-1}_a)_{\alpha\beta} d^a_{\beta} + 
\sum_{i}\frac{(V^{sp}_i-V_i)^2}{(\sigma^{sp}_i)^2}
\label{eq3}
\end{equation}
Here $\{d^a_1,d^a_2\} = \{d^a_{R\phi},d^a_{Rz}\}$ is the 
2-dimensional vector of IP components for each track $a$
entering in the PV fit and $S_a$ is the covariance matrix of
the measured quantities $\{\varepsilon^a_{R\phi},\varepsilon^a_{Rz}\}$;
since measurements in the $R\phi$ and $Rz$ planes are made independently,
the matrix $S_a$ is almost diagonal. 
$V^{sp}_i$ and $\sigma^{sp}_i$ are the beam-spot position and size
for the $x$ and $y$ coordinates. The first summation in 
equation (\ref{eq3}) runs over all tracks
selected for the PV fit. Because of our definitions (\ref{eq1}-\ref{eq2}) of 
the IP components,
the dependence of $\chi^2$ on the vertex position $\overrightarrow{V}$ is
quadratic and hence the minimisation of (\ref{eq3}) can be performed analytically.

An important part of the PV reconstruction is the selection of tracks 
and the rejection of bad measurements. Tracks with wrong hit associations 
in the vertex detector, as well as those coming from  decays 
of long-lived particles or from interactions in the detector material, bias
the fitted PV position and a special rejection procedure attempts to reduce
this bias.

For the PV computation, tracks with at least 
two $R\phi$ measurements and at least one $Rz$ measurement are selected.
First  the fit using all these tracks
($N_{tr}$) is performed and $\chi^2(N_{tr})$ is computed. After that 
each track $i$ is consecutively
removed and the corresponding $\chi^2_i(N_{tr}-1)$ is obtained.
The track $i$ giving the maximal difference 
$\chi^2(N_{tr}) - \chi^2_i(N_{tr}-1)$ is excluded  from the fit
if this difference exceeds a threshold value $\Delta$, which was set to 6.
This procedure is repeated while there are tracks 
with a $\chi^2$ difference exceeding $\Delta$. Since the beam-spot
position constraint is used for the PV computation, all tracks may be 
rejected for some events. In this case the PV coincides 
with the beam-spot and its covariance matrix corresponds to the 
beam-spot size. 
The fraction of such events is about 1\% for $Z$ hadronic events.

This fitting procedure gives an average precision of the PV 
position for $q\bar{q}$ (where $q=uds$), $c\bar{c}$, $b\bar{b}$ 
simulated $Z$ hadronic decays
of $\sigma_x = 36, 44, 60\  \mu$m 
and $\sigma_z = 43, 50, 70\ \mu$m respectively, 
although the actual precision  depends strongly on the number of tracks. 
The somewhat degraded precision for $b\bar{b}$ events is 
explained by the smaller multiplicity of primary tracks and 
by tracks from $b$-hadron decay occasionally included in the primary
vertex.

\subsection{Error and Sign of Impact Parameter}

\label{sec13}

Since the PV position is used in the definition of an IP,
the impact parameters of all tracks included in the PV fit are
correlated with each other; their correlation coefficient is about 0.2.
From equations (\ref{eq1}-\ref{eq3}) and the
standard error propagation formalism, the error on the $R\phi$ IP is given by:

\begin{equation}
\sigma^2_{R\phi} = 
\left\{ 
  \begin{array}{ll} 
    (\sigma^{tr}_{R\phi})^2 - (\sigma^{pv}_{R\phi})^2 & 
    \mbox{if the track is included in the PV fit} \\
    (\sigma^{tr}_{R\phi})^2 + (\sigma^{pv}_{R\phi})^2 & 
    \mbox{otherwise}
  \end{array}
\right.
\label{eqerr}
\end{equation}
with similar equations for 
$\sigma^2_{Rz}$.
Here $\sigma^{tr}_{R\phi}$ ($\sigma^{tr}_{Rz}$) is the error on 
$\varepsilon_{R\phi}$ ($\varepsilon_{Rz}$) coming 
from the track fit and $\sigma^{pv}$ is the error from the PV fit,
and includes implicitly the influence of all other impact parameters.
More explicitly:
\begin{eqnarray}
(\sigma^{pv}_{R\phi})^2 & = & \sum_{i,j} e_i S^V_{ij} e_j \\
(\sigma^{pv}_{Rz})^2 & = & \sum_{i,j} l_i S^V_{ij} l_j 
\end{eqnarray}
where $S^V_{ij}$ is the covariance matrix of the primary vertex fit, 
$\overrightarrow{e}$ and $\overrightarrow{l}$ are defined in 
Section \ref{sec11}, and repeated indices imply summation. The simplicity 
of the final equations is a consequence of our choice of IP components.

Using equations (\ref{eq1}, \ref{eq2} and \ref{eqerr}) the track 
significances $S_{R\phi}$ and $S_{Rz}$ are defined simply as:
\begin{eqnarray}
         S_{R\phi} & = & d_{R\phi} / \sigma_{R\phi}  \\
         S_{Rz} & = & d_{Rz} / \sigma_{Rz}  
\end{eqnarray}
The track significance thus compares the measured value of the IP
with its expected precision. This quantity 
is used as an input variable for the lifetime tagging.
Tracks from decays of long-lived particles ($\tau$'s, $b$-, $c$- and
$s$-hadrons) often have large  
IPs, significantly exceeding $\sigma_{R\phi}$ and  $\sigma_{Rz}$.

\begin{figure}
  \epsfig{figure=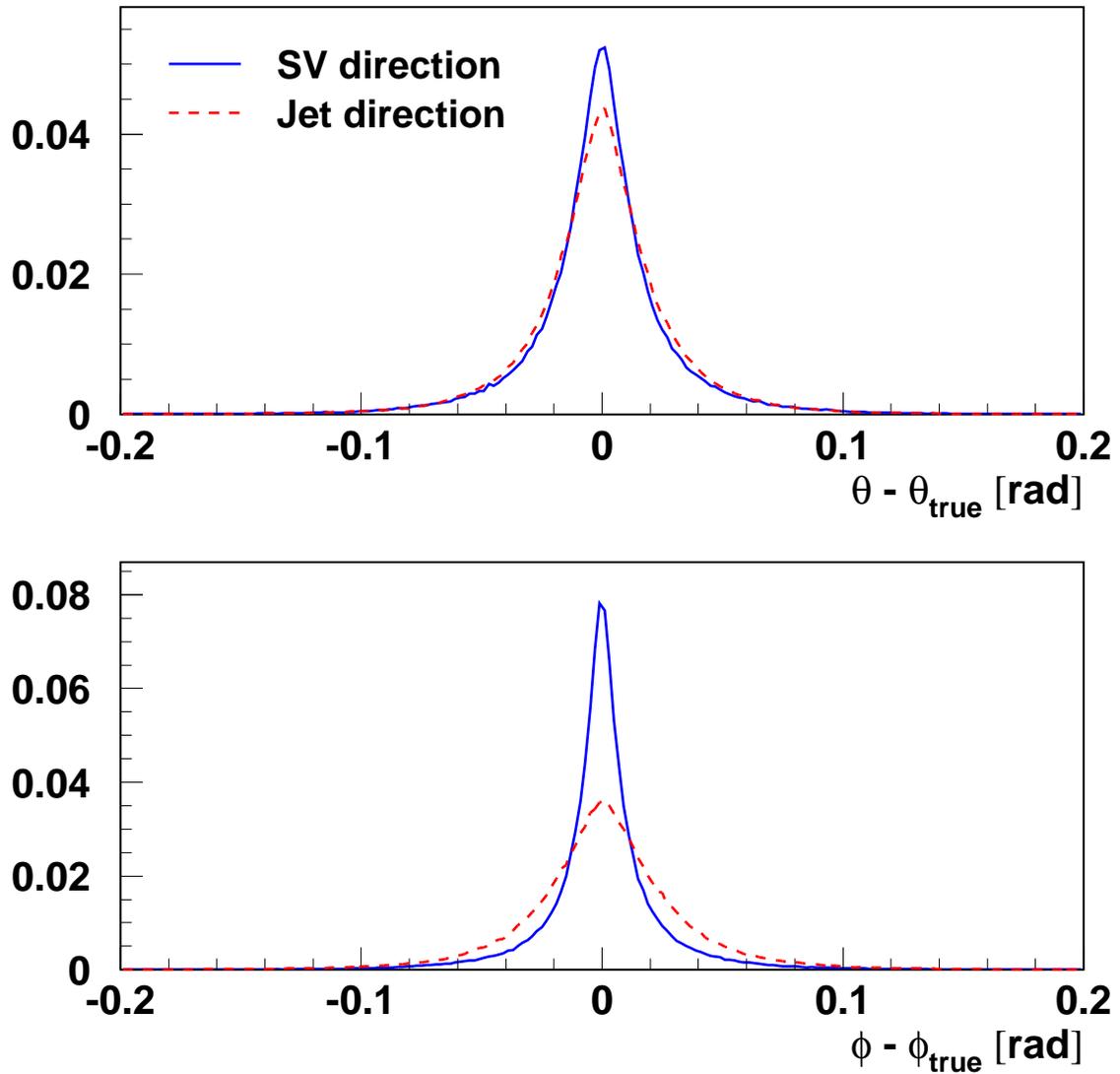,width=17cm}
  \caption{Distributions of the
    difference between reconstructed and generated directions of $b$-hadrons
    for events with a reconstructed secondary vertex, for simulated $Z$ events.
    The $B$ direction
    is defined as the direction from the primary to the secondary vertex
    (solid line) or as the jet direction (dashed line). Especially in $\phi$,
    the definition using the secondary vertex gives a better description of the
    $B$ direction.}
  \label{fig:bdir}
\end{figure}

Equations (\ref{eq1}-\ref{eq2}) define the magnitude of IP components
and their {\it geometrical} sign, while in the $b$-tagging method 
and throughout this paper the {\it lifetime} sign for IP is used. It requires 
knowledge of the flight path of the long-lived particle. In the simplest 
case the flight path is approximated by the direction of the 
jet\footnote{The default jet
clustering algorithm is JADE, with $y_{cut}$ set at 0.01. However, the user of 
the $b$-tagging package has the option of using any jet algorithm.}  
to which the given particle belongs.
Often the decay point of the long-lived particle can be reconstructed
(see Section \ref{sec22}); in this case the flight direction is defined
as the direction from the primary to the secondary vertex. 
As can be seen in fig.~\ref{fig:bdir},
this improves the measurement of the flight direction. The 
azimuthal angle precision becomes slightly better than that for 
the polar angle because the vertex detector is more precise in 
the $R\phi$ plane. To obtain the {\it lifetime} sign of the 
$R\phi$ and $Rz$ IPs, the point of closest approach in space
of the track to the estimated $B$-flight path is computed
and the sign is set negative (positive)  
if this point is upstream (downstream) of the PV position.
The significance is assigned the same sign as the IP.

With this definition, tracks from decays of long-lived particles have 
predominantly positive signs while
tracks coming directly 
from the PV are equally likely to be positive or negative.
For $b$-tagging, tracks with  positive  IP are used, thus
reducing by half the number of background tracks.

\begin{figure}
  \epsfig{figure=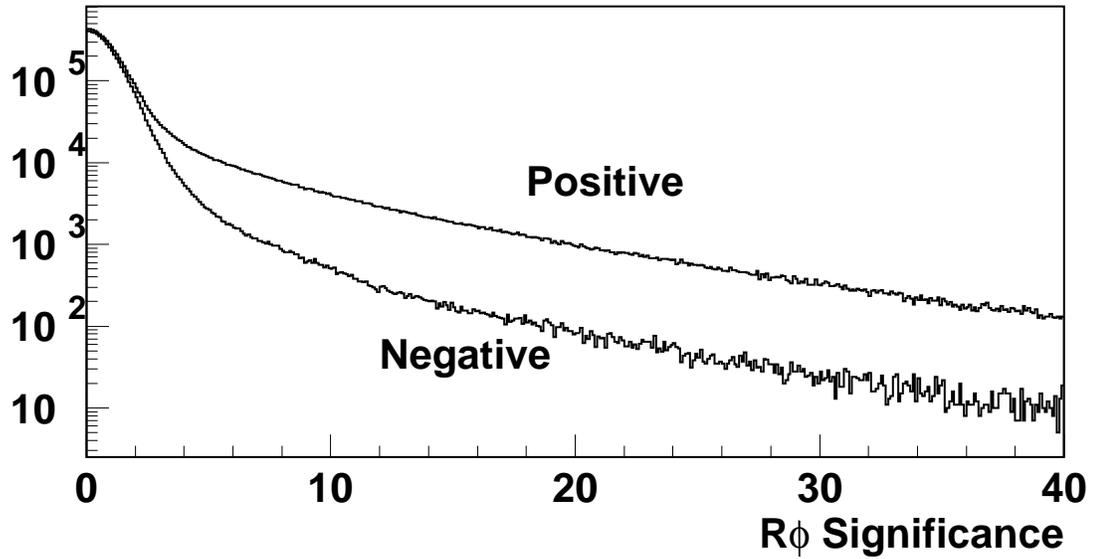,width=17cm}
  \caption{Distributions of positive and negative $R\phi$ significances
	in data $Z$ hadronic decays. The
         excess of tracks with large positive significance is due to long-lived
         particles.}
  \label{fig2}
\end{figure}

The distributions of positive and negative $R\phi$ significance  
are shown in fig.~\ref{fig2}. The excess of positively signed 
tracks with large significance is clearly seen.

\subsection{Track Probability}

\label{sec14}

The distribution of the negative track significance is determined mainly
by tracks coming from the PV, including scatters
in the detector material, tracks with wrong hit association etc, while
the contribution of tracks coming from decays of long-lived particles
is about 1\%. 
This distribution can thus be used to define the
probability $P(S^0)$ for a track from the PV to have the  measured 
value of the modulus of its significance  exceeding the value $S^0$. 
This function is obtained by integration of the probability density
function of the negative 
significance $f(S)$ from $S^0$ to infinity and assuming that $P(S^0)$ 
is the same for primary tracks with either positive or negative significance:

\begin{equation}
P(S^0) = \int_{S^0}^{\infty} f(S) dS 
\end{equation}

\begin{figure}
  \epsfig{figure=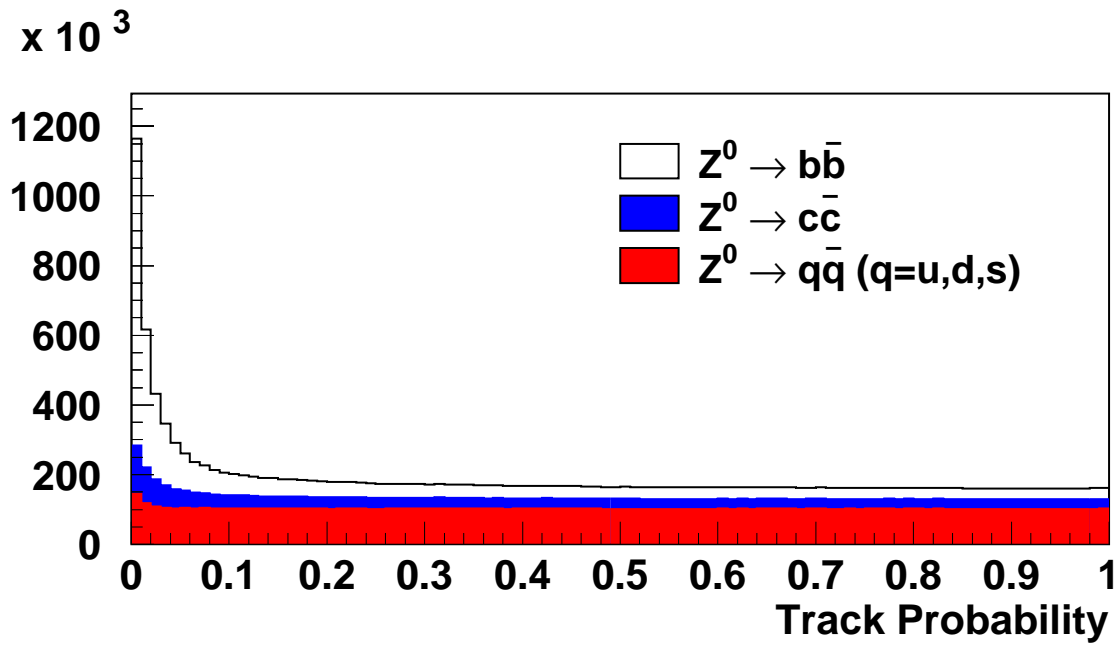,width=17cm}
  \caption{Simulated distributions at the $Z$ of the $R\phi$ track IP 
    probabilities
    for different quark flavours, for tracks with positive lifetime-sign IPs.}
  \label{fig3}
\end{figure}

By definition, tracks from the PV should have a flat
distribution of $P(S^0)$ between 0 and 1, while tracks from
decays of long-lived particles and with large positive values of $S^0$ 
have small values of $P(S^0)$, reflecting the small probability
for tracks from the primary vertex to have such large values of the IP and
hence of $S^0$. As an example, fig.~\ref{fig3} shows 
the distribution of $P(S^0_{R\phi})$ for tracks with positive IP.
The peak at small values of $P(S^0_{R\phi})$ is produced mainly
by the long-lived particles. The transformation from significance 
to track probability is referred to as the calibration
of the detector resolution.

For LEP1 analyses, the above calibration was performed using tracks from $Z$
decays. At LEP2, there was the possibility of again using $Z$ data for
calibration; each year, short runs at the $Z$ were taken before the start of
and also interspersed with
the high energy running. Alternatively, the calibration 
could be carried out using
the same type of data as used to perform the relevant physics analysis. Thus
for the Higgs search of Section \ref{Higgs}, 4-jet events were also used for
calibration in the channel where the $H$ and $Z$ both decay to 2 jets, 
while in the corresponding channel where the $Z$ decays to two
neutrinos, calibration was performed using `2-jets + missing energy' events.
The use of calibration samples closely related to the data sample in principle
allows for the following possible effects:

\begin{itemize}
\item
because of possible movements of the relative positions of the 
different parts of the vertex
detector with respect to the rest of DELPHI, the calibration could be
time-dependent;
\item
track confusion and lifetime-signing (and hence calibration) could depend
on the event topology;
\item
the IP resolution changes with polar angle $\theta$. The jet distribution in
$\theta$ depends on the particular physical process considered;
\item
the  IP resolution is also energy dependent. The jet and track energy spectra 
depend on the physical process.
\end{itemize}

The calibration was performed separately for categories of tracks 
with different lifetime sensitivities. The categories were
determined by the number of associated VD hits. A small number of VD hits 
associated with a track is often caused by incorrect reconstruction, 
and the significance distribution of such tracks has a larger non-Gaussian 
tail.
By using a different track probability for them,
this difference was taken into account. 
%
This approach was also used for the analysis of the data collected in 2000, 
when one out of the 12 sectors
of the TPC  was not operational during the last part of the data taking; 
tracks reconstructed without the TPC have worse precision, which requires
using a separate track probability  for them.

Another property of the track probability is that it 
can be defined directly from the data. This is very important, in that 
it allows the calibration of the detector resolution independently of the 
simulation. Such calibration allows to take into account possible 
differences between data and simulation.
As a consequence, it also reduces the systematics due to detector effects in 
physics measurements.

For the construction of $P(S^0)$ it is important to reduce
the contribution of tracks coming from the decay of long-lived particles.
Using the negative significance distribution partially solves this
problem. Additional suppression of the lifetime information is achieved
by applying anti-$b$ tagging to the event sample used for calibration.
This anti-$b$ tagging is based on tracks with positive IPs, and hence does not
bias the negative significance distribution.
The anti-$b$ tagging reduces the fraction of $b\bar{b}$ events 
in the selected sample of hadronic $Z$ events from 21.6\% to less 
than 5\% and the contribution of tracks from decays of $b$-hadrons 
is reduced correspondingly. Additional selection criteria for tracks
used for calibration decrease the contribution from the decay products 
of light long-lived hadrons ($K^0$, hyperons) and hence reduce the tail of the 
significance distribution. They are specified in the next section.


\subsection{Lifetime Probability}
\label{sec15}

Track probabilities are directly used
to construct a lifetime probability~\cite{aleph}. For any group
of N tracks it is defined as:
\begin{equation}
P_N = \Pi \cdot \sum_{j=0}^{N_{R\phi}+N_{Rz}-1} (-\log \Pi)^j/j!,~~
\mbox{where}~~
\Pi = \prod_{i=1}^{N_{R\phi}} P(S_{R\phi}^i) \cdot
\prod_{i=1}^{N_{Rz}} P(S_{Rz}^i) 
\label{eqpn}
\end{equation}
Here $P(S_{R\phi}^i)$, $P(S_{Rz}^i)$ are the track probabilities
and $N_{R\phi}$, $N_{Rz}$ are the number of $R\phi$ and $Rz$ IPs
used in the tagging.
The definition of
$P_N$ thus ignores the small off-diagonal elements of the IP error matrix
and the correlation between different IPs
coming from the use of the common PV. 

The variable $P_N$ has a simple and straightforward definition 
and can be computed for any group of tracks (e.g. a jet, hemisphere 
or whole event) which makes it flexible and easily adjustable
to different physics applications. No other combination of IP measurements
was found to give a better selection of $b$-quarks.

An attractive feature of lifetime tagging is that it is constructed
using only the track IPs. This provides the possibility 
of achieving a good description of the $b$-tagging efficiency by 
the accurate tuning of the track resolution in simulation,
as described in Section \ref{tuning}. It allows a significant decrease in
the systematic uncertainties due to detector effects in physics measurements.

The meaning of the variable $P_N$ is very similar to that of track
probability $P(S)$: it is the probability for $N$ tracks coming from the PV
to have the product of their track probabilities exceeding the 
observed value. It varies between 0 and 1 and has a flat distribution for 
any group of $N$ uncorrelated tracks coming from the PV. 
The contribution of tracks from secondary decays shifts 
$P_N$ to lower values, producing a peak near 0.

\begin{figure}
  \epsfig{figure=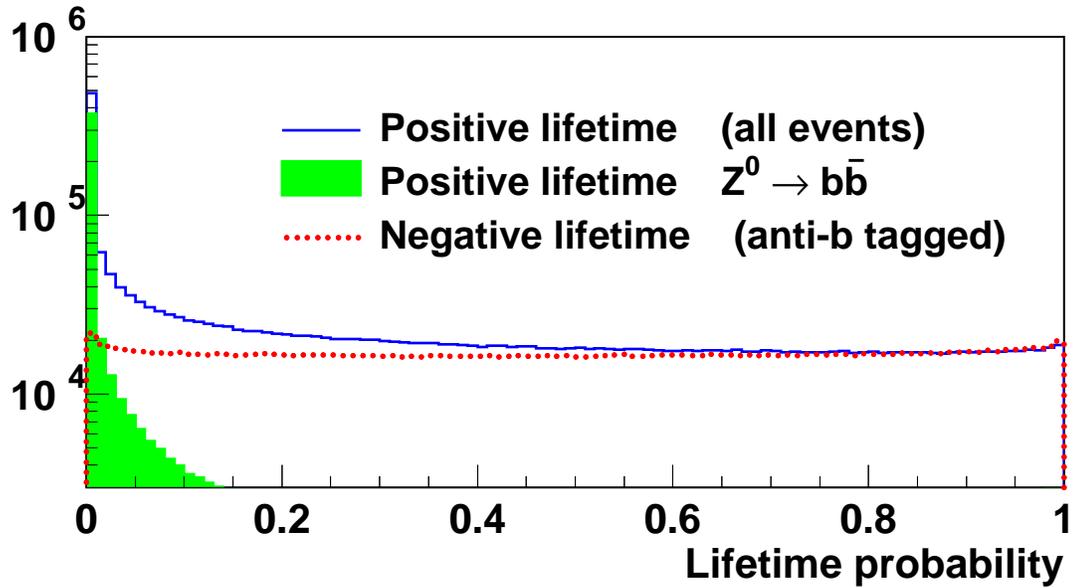,width=17cm}
  \caption{Distributions of positive and negative lifetime probabilities 
	($P^+_N$ and $P^-_N$ respectively) for
    simulated $Z$ hadronic events.}
  \label{bt-ltp}
\end{figure}

The flat distribution of $P_N$ for primary tracks can be
verified by computing $P_N^-$ for the sample of all tracks with negative 
impact parameters in anti-$b$ tagged events. 
As explained in Section \ref{sec14},
the contribution of tracks from decays of long-lived particles in such a
sample is small. The distribution of $P_N^-$ is shown as the dotted curve in 
fig.~\ref{bt-ltp}. It is relatively flat, although there is a small peak 
near zero. This peak is produced by tracks from decays of 
long-lived particles, 
which are occasionally assigned to have negative IPs because of the 
error in their flight direction estimate. However, the value of this
excess is significantly less than the peak in the distribution
of $P_N^+$, computed using positive IPs. This distribution is shown
as the solid curve in fig.~\ref{bt-ltp}. The latter peak
is mainly produced by the $b \bar{b}$ events, as can be seen from 
fig.~\ref{bt-ltp}.

The separation of tracks into two samples depending on the 
sign of their IP is very important. The 
sample of negative IP tracks is used for the calibration of the detector 
resolution and the quality of this calibration is verified by the $P^-_N$ 
distribution. In contrast, positive IP tracks are used for $b$-tagging. 
Thus the samples of tracks used in the calibration and in the analysis do not
overlap.
The $P^-_N$ distribution also gives a good estimate of the background level 
from light quarks at the corresponding value of $P^+_N$.

The $R\phi$ and $Rz$ components enter in eqn. (\ref{eqpn}) separately.
As explained in Section \ref{sec11}, the fraction of wrong
measurements in $Rz$ is higher. Therefore, tighter selection criteria
are applied to tracks for the $Rz$ IP, and for some tracks
only the $R\phi$ measurement is used.  
Thus, the separate treatment of IP components also allows
the use of information which would otherwise  be lost.

More specifically, the conditions applied to the tracks are as follows.
All tracks with positive IP and at least one measurement
in the VD are candidates for lifetime tagging. Tracks coming 
from reconstructed $K^0$ or $\Lambda$ 
decays\footnote{For $V^0$ reconstruction procedure see \cite{delphi}. }
are rejected. Both $R\phi$ and $Rz$ IP components are required 
to be less than 0.2 cm, although this condition is removed if the
track comes from a reconstructed secondary vertex (see Section \ref{sec22}).

One more parameter is used to provide additional suppression of
bad $Rz$ measurements.
It is the distance $D$ of closest approach in 3-dimensions  between the track 
and the expected flight path of the long-lived particle, defined in the 
section \ref{sec13}. All tracks, both from the 
primary and secondary vertices, should have a small value of $D$ provided
the secondary vertex is close to the estimated flight direction.
Therefore a large value of $D$ with respect to its expected precision
$\sigma_D$ is used to identify wrong IP measurements. $Rz$ IP measurements 
are rejected if $D/\sigma_D$ exceeds 2.5. Both $R\phi$ and $Rz$ measurements
are excluded from lifetime tagging if $D/\sigma_D$ exceeds 10.

\begin{figure}
  \epsfig{figure=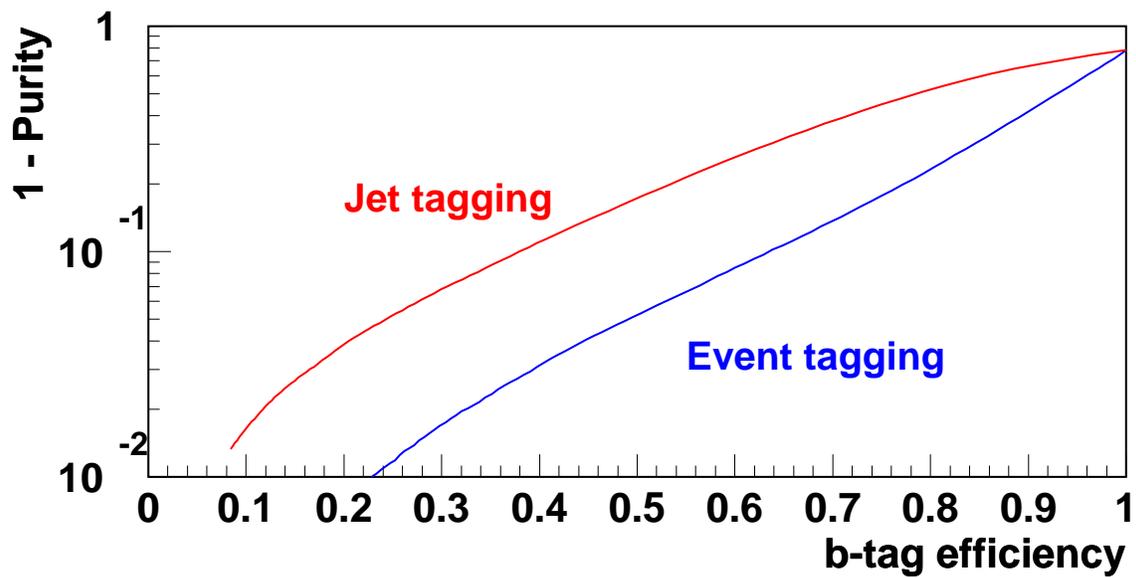,width=17cm}
  \caption{Background suppression in simulated $Z$ hadronic events using
    lifetime tagging. The efficiency is the fraction of $b$-jets that are
tagged as coming from $b$-quarks, while the purity is the fraction of tagged
jets that are really from $b$-quarks.}
  \label{bt-ltt}
\end{figure}

Figure~\ref{bt-ltt} shows the performance of the lifetime probability applied
to simulated hadronic decays of the $Z$. The figure shows the efficiency of 
the $b$-quark selection versus the contamination of the selected
sample by ($u,d,s,c$) flavours ($N_{udsc}/(N_{udsc} + N_b)$).
The suppression of background
flavours is shown for tagging of one jet (i.e. using only 
tracks from the given jet), and for the whole event. Event tagging is more 
efficient because $b$-quarks are produced in pairs. 

$b$-tagging 
using only lifetime information is rather efficient and is sufficient 
for the needs of many physics applications. An important feature is its
simple control of the tagging efficiency in 
simulation. Its performance is however substantially enhanced 
by including additional discriminating variables, such as the mass at the 
secondary vertex or the presence of energetic leptons. This method 
of `combined $b$-tagging' is described in Section \ref{sec3}.



\subsection{Tuning}
\label{tuning}

Almost all precision measurements and searches for rare processes 
rely on a comparison of the observed
data distributions with those predicted by a detailed 
simulation. For this comparison both the generation of the intrinsic
physical processes and the simulation of detector response must be as
realistic as possible.
For the selection of events containing $b$-hadrons, 
the most important variables are the track 
IPs, therefore 
the description of the IP resolution can significantly influence the 
physics result and the value of the systematic uncertainty.

The generated events in the DELPHI experiment are processed by the 
detector simulation package \cite{DELSIM} and the same reconstruction program 
\cite{DELANA} as for the data. For the simulation, the reconstruction program
first applies some additional smearing to the reconstruction inputs to improve
agreement with the particular data set being represented. For the vertex
detector, this includes 
applying corrections for inefficient regions, adding noise hits, and randomly
modifying the positions of the modules to simulate the effects of 
residual misalignments in the real data. 

\begin{figure}[htb]
\begin{center}
   \epsfig{file=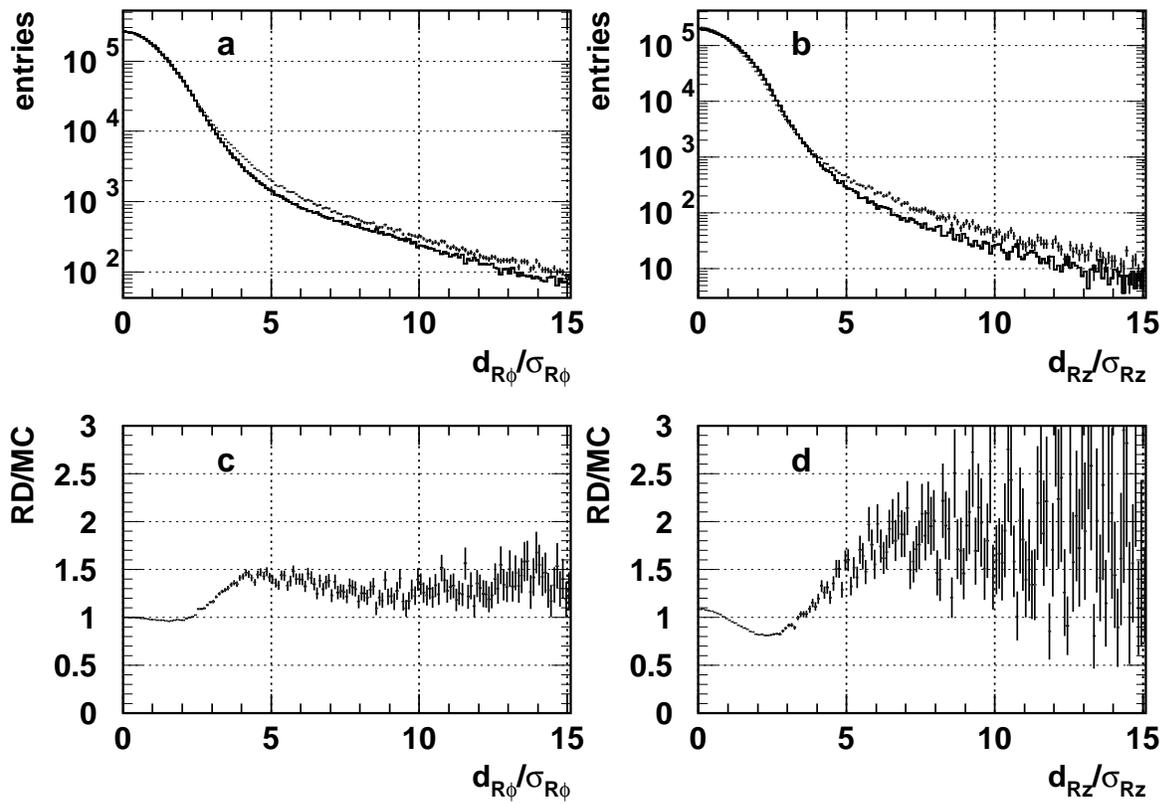,width=17.cm}
\end{center}
\caption{{\bf a)} and {\bf b)} The $R\phi$ and $Rz$ significance 
distributions for tracks with negative IP.
The points with errors are real data, the histogram is simulation. 
{\bf c)} and {\bf d)} The ratios of these distributions (data divided by
simulation). }
\label{tun:fig2}
\end{figure}

However, even after this procedure 
some disagreement between data and simulation in the track resolution 
description remains. 
This difference can be clearly seen, for example, 
in the distribution of the track significance (see fig. \ref{tun:fig2}).
Any disagreement in this quantity can result in a large 
discrepancy in the $b$-tagging description.

A detailed description of the 
method including  the correction of the detector resolution 
in the $R\phi$ plane
for the initial micro-vertex detector \cite{nim1} is given in
\cite{nim_gc}. The application of this method for the tuning in the $Rz$ 
plane for the DSVD at LEP1 and the SiT at LEP2
is similar and consists of the following steps:

\begin{itemize}
\item  the appropriate parameterisations of the negative lifetime-signed
IP ($d_{R\phi}$ and $d_{Rz}$) distributions are determined;
\item  the numerical coefficients for these parameterisations are extracted 
from the data;
\item
the errors of $d_{R\phi}$ and $d_{Rz}$ given by the track fit 
are corrected both in data and in the simulation according to the 
parametrisation obtained while the correlation
between $d_{R\phi}$ and $d_{Rz}$ is not changed;
\item additional smearing of $R\phi$ and $Rz$ IPs in simulation
is performed in order to reproduce the observed real data distributions.
\end{itemize}

\begin{figure}[htb]
\begin{center}
   \epsfig{file=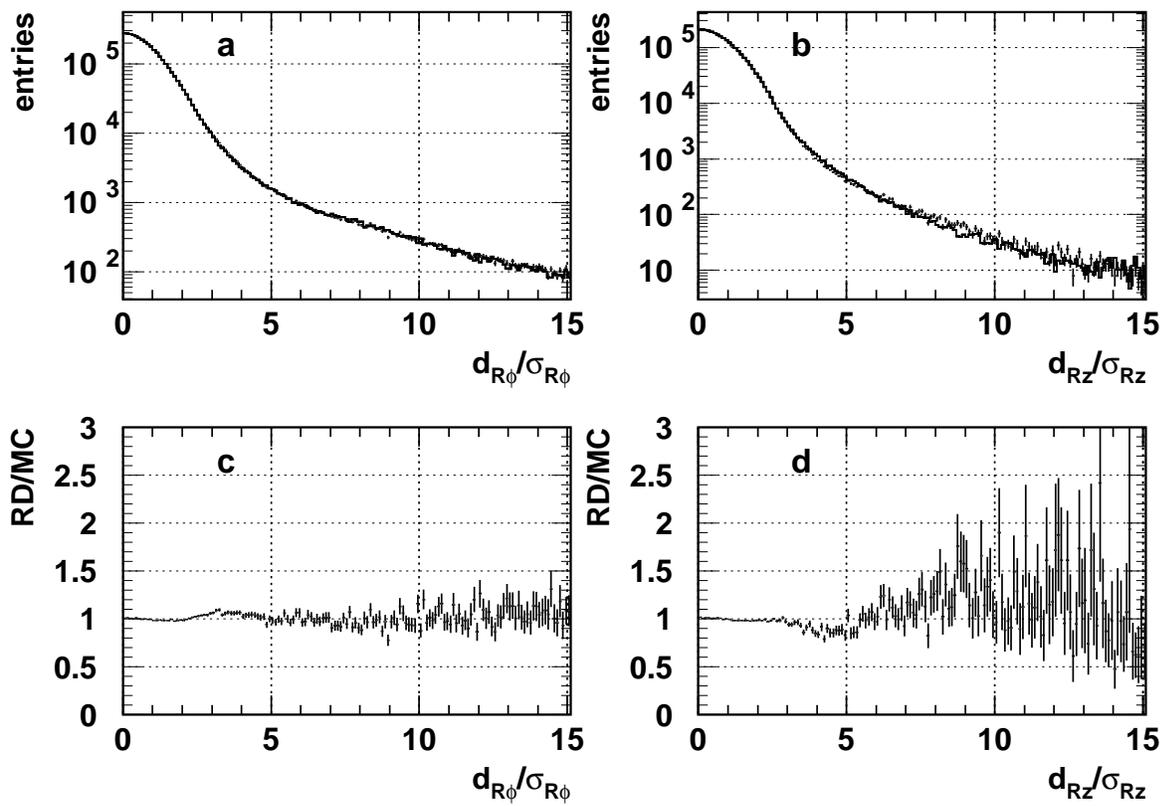,width=17.cm}
\end{center}
\caption{{\bf a)} and {\bf b)} The $R\phi$ and $Rz$ significance 
distributions for tracks with negative IP after the tuning procedure.
The points with errors are real data, the histogram
is simulation. {\bf c)} and {\bf d)} The ratios of these distributions 
(data divided by simulation). The improvement resulting from 
tuning is clearly visible.}

\label{tun:fig4}
\end{figure}

\begin{figure}[htb]
\begin{center}
\mbox{\epsfig{file=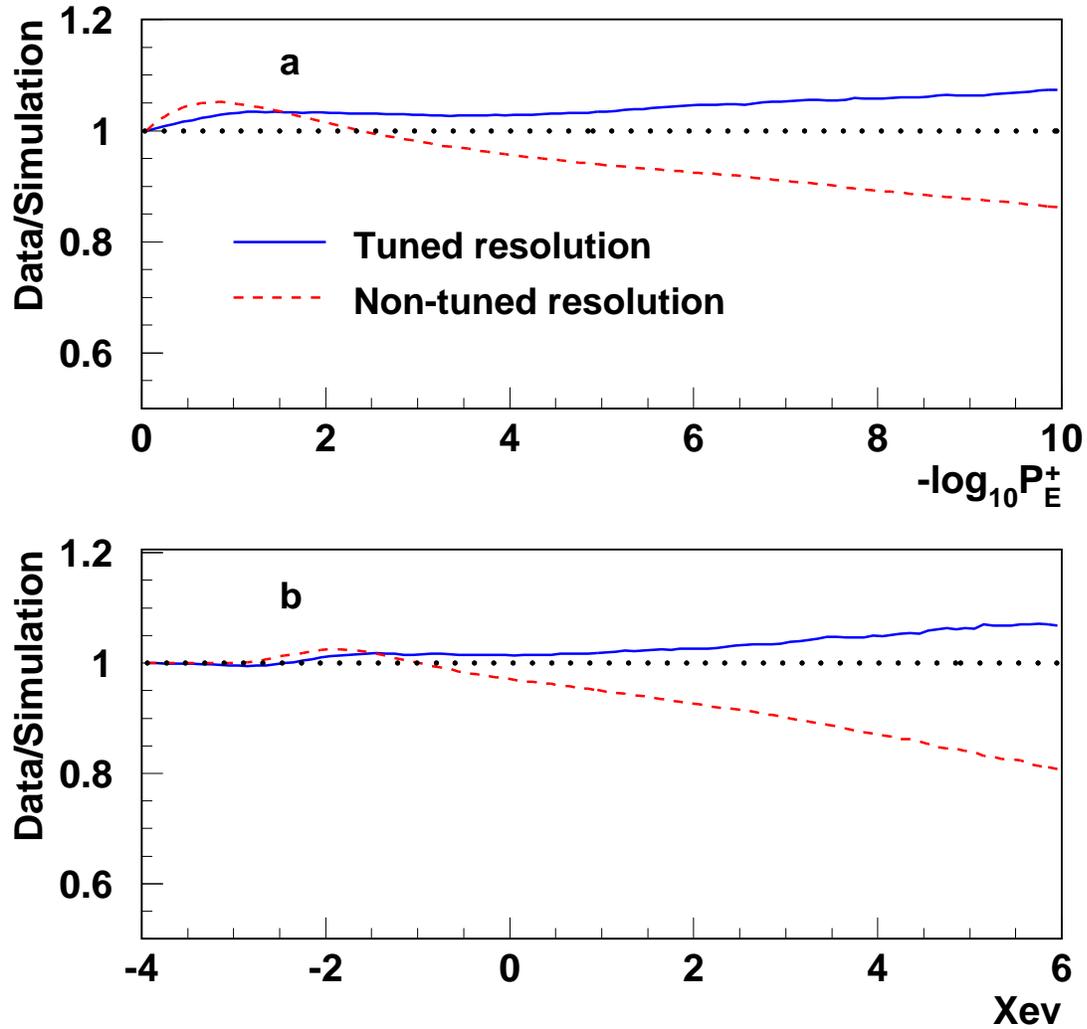,width=16cm}}
\end{center}
\caption{
The integrated data to simulation ratio of 
the fraction of selected hemispheres  as a function of 
the cut on the $b$-tagging variable for {\bf a)} the lifetime tagging 
variable $P^+_E$, calculated using all positive lifetime IP tracks
in an event (see Section \ref{sec15}) and {\bf b)} 
combined tagging variable $X_{ev}$(see Section \ref{sec3}), 
with tuned (full line) and non-tuned (dashed line) track resolution.}
\label{tun:fig3}
\end{figure}

The improvements in the significance 
description after applying this method 
can be seen in fig. \ref{tun:fig4}.
Fig. \ref{tun:fig3} shows the 
data to simulation ratio of the selection efficiency as a function 
of the cut on the $b$-tagging variables.
For the non-tuned version of $b$-tagging (dashed line)
the difference between data and simulation is very significant
for strong $b$-tagging cuts, 
corresponding to purer  samples of $B$ events.
The tuning  results in  better agreement for 
both the lifetime and the combined $b$-tagging variables, the latter being 
described in the next section. 
The remaining differences between data and simulation can be explained by the
uncertainties of the modelling of $B$ decay and to a lesser extent its hadronic
production.

This tuning procedure is incorporated in the $b$-tagging 
package and is used in all DELPHI measurements involving $b$-quark selection.

\section{Combined Tagging}
\label{sec3}

Efficient utilisation of different properties of $b$-hadrons
requires the development of a technique for their combination 
into a single tagging variable. The simplest solution of applying 
some system of cuts on different discriminating variables, which was tried 
in other collaborations\cite{alrb,sld}, is not optimal due 
to a significant overlap between the signal and background for some of them. 
Instead, DELPHI uses a likelihood ratio method of variable 
combination\cite{comb,gbb}. This 
approach has the important advantage of being  technically very simple while 
at the same time providing powerful separation of signal and background.
For independent variables, it gives optimal tagging, i.e. the best possible 
background suppression for a given signal efficiency \cite{anderson}.
It can easily be extended to any number of discriminating
variables, and can deal with different numbers of variables in different
events. However, its practical application requires the careful
selection of variables with reduced correlations among them.
The description of this likelihood ratio method, the set of variables used
and the performance of the DELPHI combined $b$-tagging is given below.

\subsection{Description of Method}
\label{descrip}
The combined tagging variable $y$ in the likelihood ratio method is defined as:

\begin{equation}
\label{eq-like}
y=\frac{f^{bgd}(x_1,...,x_n)}{f^{sig}(x_1,...,x_n)}
\end{equation}
where $f^{bgd}(x_1,...,x_n)$, $f^{sig}(x_1,...,x_n)$ are the probability
density functions of the discriminating variables $x_1,...,x_n$ for
the background and the signal respectively. The selection
of all events with $y < y_0$ gives the optimal tagging of the signal.
It should be stressed that such tagging is absolutely the best
for a given set $x_1,...,x_n$ of variables.

In practical applications the determination and utilisation of
multi-dimensional probability density functions is quite difficult for $n>2$. 
The solution consists in a special selection 
of discriminating variables having reduced correlations among them.
In the limit of independent variables\footnote{Two variables are independent if,
for the signal and for each separately treated background component (e.g. $c$
and $uds$), the
distribution of one is independent of any selection on the other.}, 
expression (\ref{eq-like}) becomes:
\begin{eqnarray}
\label{eq-comb1}
y & = & \prod_{i=1}^{n} \frac{f^{bgd}_i(x_i)}{f^{sig}_i(x_i)} = 
\prod_{i=1}^{n} y_i; \\
\label{eq-comb2}
 y_i & = & f^{bgd}_i(x_i)/f^{sig}_i(x_i)
\end{eqnarray}
where $f^{bgd}_i(x_i)$, $f^{sig}_i(x_i)$ are probability density functions
of each individual variable $x_i$ for the background and signal,
and are determined from simulation.

This scheme is used in DELPHI
to construct the combined $b$-tagging. For each individual variable $x_i$
the value $y_i$ is computed from (\ref{eq-comb2}); 
the combined tag $y$ is defined as the product of the $y_i$. 
It is not exactly optimal any more,
because the discriminating variables are not independent,
but the variables are chosen such that the correlations between them
are small enough that the resulting tagging is very close to optimal.
Furthermore, the efficiencies and mis-tag rates are determined from
simulation (and sometimes from the actual data), thereby taking into account 
any small correlations. 

In DELPHI all discriminating variables and the $b$-tagging itself
are computed independently for each jet in an event, where ideally all 
tracks coming from the fragmentation of the $b$-quark and from the decay of 
the $b$-hadron are combined in one jet by a jet clustering algorithm. 
In this case the background for the $b$ quark selection can be separated 
into two different parts -- jets generated by  $c$-quarks and by light 
($q = u,d,s$) quarks. These two components are independent and have very 
different distributions of discriminating variables. 



To define the extra discriminating variables for the $b$-tagging,
tracks are selected so as to come preferentially from $b$-hadron decay.
For this purpose all jets in an event are classified
into 3 categories. In the first category all jets with one or more reconstructed
secondary vertices are included. 
A reconstructed secondary vertex
provides a clean selection of $b$-hadron decay products and a large 
number of discriminating variables can be defined in this case.
If the secondary vertex is not reconstructed, tracks from the $B$ 
decay are selected by requiring the track significance probability to be
less than 0.05,
and the second category includes all jets with at least 2 such offsets.
This criterion is less strong, allowing more background jets to pass 
the cut. Finally, if the number of offsets is less than 2, 
the jet is included in the third category and in this case 
only a reduced set of inclusive discriminating variables, like the 
lifetime probability (see Section \ref{sec15}), is used.
In $Z$ hadronic events the fractions of jets classified
into categories 1,~2,~3 are 44\%,~14\%,~42\% respectively for  $b$-quark,
8\%,~8\%,~84\% for $c$-quark and 0.6\%,~2.8\%,~96.6\% for light
quark jets.


The tagging variable $y_\alpha$ for a jet of category $\alpha$ is 
defined as:

\begin{eqnarray}
\label{eq-comb}
y_{\alpha}~~~ & = & n^c_\alpha/n^b_\alpha \prod_{i} y_{i,\alpha}^c + 
n^q_\alpha/n^b_\alpha \prod_{i} y_{i,\alpha}^q; \\
y_{i,\alpha}^{(c,q)} & = & f_{i,\alpha}^{(c,q)}(x_i) / f_{i,\alpha}^b(x_i) 
\nonumber
\end{eqnarray}
where $f_{i,\alpha}^q(x_i)$, $f_{i,\alpha}^c(x_i)$, $f_{i,\alpha}^b(x_i)$ 
are the probability density functions of $x_i$ in jet category $\alpha$ 
generated by $uds$, $c$ and $b$ quarks respectively and 
$n^q_\alpha$, $n^c_\alpha$ and $n^b_\alpha$ are their normalised rates, 
such that $\sum n^q_\alpha = R_q$, $\sum n^c_\alpha = R_c$, and
$\sum n^b_\alpha = R_b$. $R_q$, $R_c$ and $R_b$ are the normalised production
rates of different flavours and $R_q + R_c + R_b = 1$.

As can be seen from eqn. (\ref{eq-comb}), the classification 
into different categories effectively works as an additional discriminating 
variable with the discrete probabilities given 
by $n^{(q,c,b)}_\alpha$. For example,  the $b$-purity of a sample of jets with 
reconstructed secondary vertices is about 85\%. However, the primary purpose 
of this separation is to allow 
the use of a larger number of discriminating variables when a secondary vertex
is found.
The search for the secondary $B$ decay vertex
is thus an important ingredient of DELPHI $b$-tagging.

It is often convenient to define $X_{jet} = -\log_{10}y_\alpha$ 
as the jet tagging
variable, and this variable is used in all applications described in Section
\ref{applications}. The event variable $X_{ev}$ 
is defined as the sum of the largest two $X_{jet}$ values for the individual 
jets in the event.

\subsection{Secondary Vertex Reconstruction}
\label{sec22}

A secondary vertex (SV) is searched for 
in each jet of the event. In the first stage all possible
combinations of pairs of tracks are selected as SV candidates
if they have a common vertex with the $\chi^2$ of the fit
less than 4. After that all tracks from the same jet are tested
one by one for inclusion in a given SV candidate. The track
producing the smallest change $\Delta$ of the vertex fit $\chi^2$ is included
in the SV candidate if this change does not exceed the threshold
$\Delta=5$. This value and all other numerical parameters of the algorithm 
were selected by optimising
the efficiency of the SV reconstruction and background suppression.
This procedure is repeated until all tracks satisfying the above condition
are included in the SV candidate. The SV candidate is rejected
if the distance to the primary vertex divided by its error is less
than 4. Additionally, at least two tracks in the SV candidate are required
to have VD measurements in both $R\phi$ and $Rz$ planes.

The decay of the $b$-hadron is usually followed by decays of one or
two $D$ mesons, thereby producing several secondary vertices.
It thus often happens that some
secondary tracks cannot be fitted to a single secondary vertex.
However, the spatial distance between any secondary track and the 
flight trajectory of the $b$-hadron should be small since the $D$-mesons 
tend to travel in the direction of the initial $b$-hadron. 
Using this property 
some tracks of far-decaying $D$ mesons can be recuperated, which is important 
for the computation of such quantities as the $b$-hadron mass.
The flight trajectory of the $b$-hadron is defined as the vector
from the primary to the secondary vertex.
Any track from the same jet having an $R\phi$ or $Rz$ component
track probability less than 0.03 and not included in the SV fit 
is attached to the SV candidate if its distance in space to the flight 
trajectory divided by its error is less than 3. Although not included
in the SV fit, such tracks are used 
in the computation of all discriminating variables.

Sometimes the decay vertex of a $D$-meson can be reconstructed separately
from the $b$-hadron decay vertex, which then results in two or more 
secondary vertices in the same jet. In that case,
all tracks included in these vertices are combined for the computation
of the $b$-tagging discriminating variables.

Three additional criteria are used to suppress the background
of light quarks in the sample of jets with secondary vertices.
The first one makes use of the momentum vector of the $b$-hadron.
This  is defined as the sum of the momenta of all tracks included
in the SV candidate. Additionally, the momenta of all other neutral and
charged particles with pseudo-rapidity exceeding 2 are also included;
the pseudo-rapidity is computed with respect to the flight direction
of the $b$-hadron. Then, the trajectory directed along the $B$-momentum 
and passing through the SV position is constructed and the 
impact parameter $\delta_{SV}$ of this trajectory with respect 
to the primary vertex is computed. For a real $b$-hadron the 
momentum direction should be close to the flight direction and $\delta_{SV}$
should be small compared to its 
error\footnote{
$\sigma_{\delta_{SV}}$ is computed using the uncertainties in the positions
of primary and secondary vertices and the uncertainty in the $B$-hadron flight
direction estimate, obtained from the simulation.}
$\sigma_{\delta_{SV}}$, while for a 
false secondary vertex the flight and momentum directions are much less 
correlated. Therefore SV candidates with 
$(\delta_{SV}/\sigma_{\delta_{SV}})^2>12$ are rejected.
For the second criterion the lifetime probability 
using all tracks included in the SV candidate is computed and the 
candidate is rejected if this probability exceeds 0.01.
The third criterion requires the distance between the primary 
and secondary vertex to
be less than 2.5 cm, because the contribution of false secondary vertices
and of strange particle decays 
becomes rather high at large distances. 
Any background jet with a
very distant SV would give an extremely strong $b$-tagging
value and this cut effectively rejects such cases.

Candidates remaining after these selections are considered
as  reconstructed secondary vertices. With this procedure
a SV is reconstructed for about 44\% of jets
with $b$-hadrons (50\% for jets inside the VD acceptance). 
The $b$-purity of the sample of jets with a reconstructed SV 
is about 85\% for hadronic decays of the $Z$, which should be compared 
with the initial $b$-purity of about $22\%$. More than one SV in 
a single jet is allowed, reflecting the possibility of cascade 
$(B\to D)$ decays giving rise to distinguishable secondary vertices. 
In this case the tracks from all secondary vertices 
are combined for the computation of the SV discriminating variables.


\subsection{Discriminating Variables}
\label{sec23}
In this section the discriminating variables
used in $b$-tagging are described. All definitions are given 
first for jets with reconstructed 
secondary vertices. Then the modifications  for other jet categories
are described. All discriminating variables, except the transverse
momentum of a lepton, are computed using the charged particles included
in the secondary vertex.

{\it The jet lifetime probability}, $P_j^+$, is constructed using equation 
(\ref{eqpn}) from the positive IPs of all tracks 
included in the jet. 

\begin{figure}
  \epsfig{figure=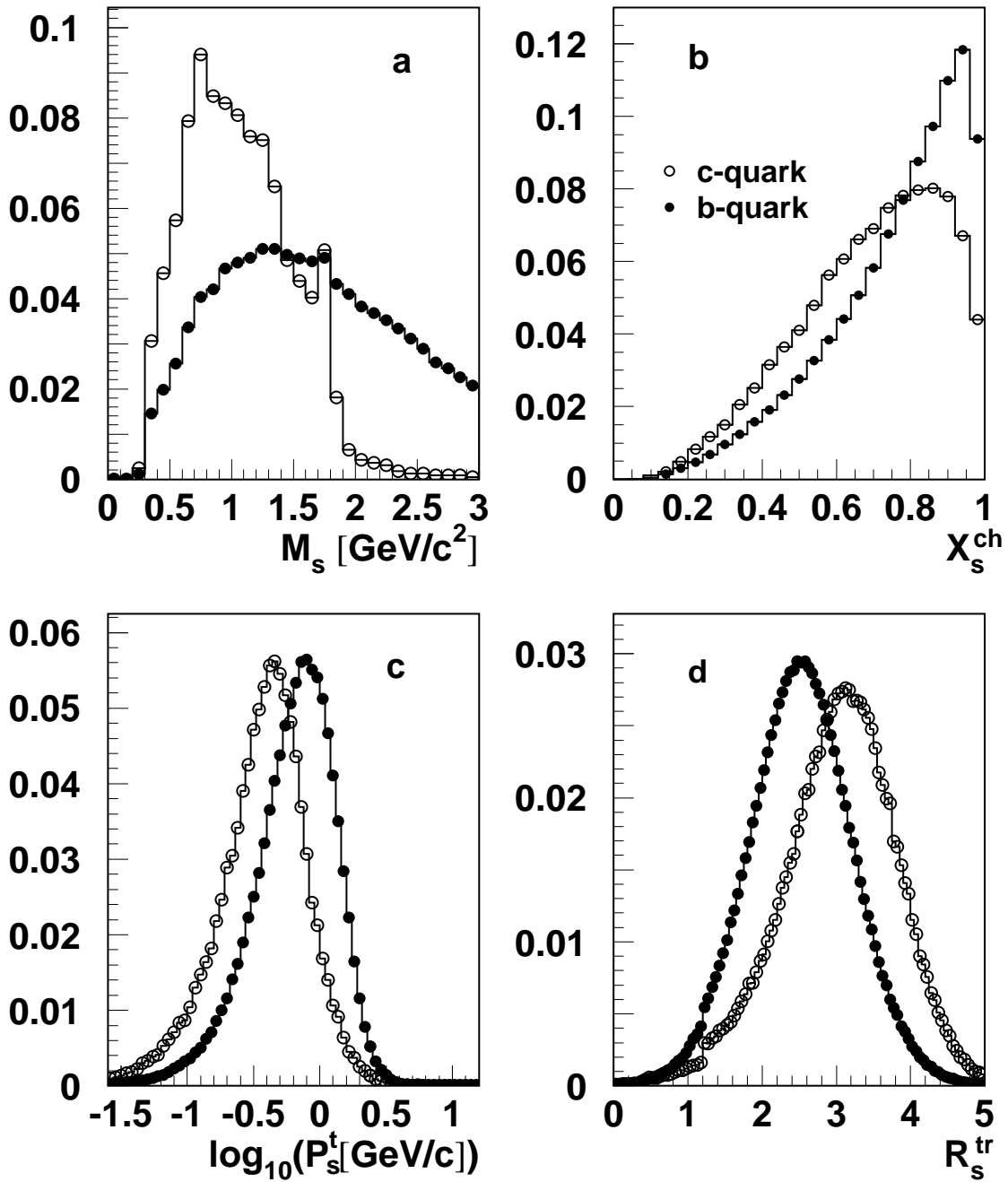,width=17cm}
  \caption{Distributions of discriminating variables for $b$- and 
    $c$- quark jets for simulated $Z$ hadronic events. 
{\bf a)} Mass of particles in SV. {\bf b)} Fraction of charged jet 
energy included in SV. {\bf c)} Transverse momentum at SV. 
{\bf d)} Rapidity for each SV track.}
  \label{bt-tv}
\end{figure}

{\it The mass of particles}\footnote{For computation of 
discriminating variables,
such as mass or track rapidity, elsewhere in this paper,
charged particles are given the pion mass, whereas neutrals (except for 
$K^0_s$ and $\Lambda$) are assumed to be massless.} 
{\it combined at the secondary vertex}, $M_s$,
is very sensitive to the quark flavour. The mass at the secondary
vertex in a jet generated by a $c$-quark is limited by the mass of
the $D$-meson, which is about 1.8 GeV/c$^2$, while the mass in 
a $b$-jet can go up to 5 GeV/c$^2$. The limit of 1.8 GeV/c$^2$ for
the $c$-jets can be clearly seen in fig.~\ref{bt-tv}(a). Some $c$-jets do have
a higher value of $M_s$ due to tracks incorrectly attached to the SV.


{\it The fraction of the charged jet energy included in the secondary
vertex}, $X_s^{ch}$, reflects the differences in the fragmentation
properties of different flavours. The fragmentation
function for the $c$-quark is softer than for the $b$-quark, as seen
in the distribution of $X_s^{ch}$ in fig.~\ref{bt-tv}(b). 

{\it The transverse momentum at the secondary vertex}, $P^t_s$, 
first introduced
by the SLD collaboration\cite{sld}, takes into account missing particles
not included in the SV definition. $P^t_s$ is defined as the 
resultant transverse momentum (with respect to the $b$-hadron's estimated 
flight direction) of all charged particles attached to the SV.
Missing particles can be neutrinos from semileptonic decay, other neutral
particles or non-reconstructed charged particles. In all cases, due to 
the high mass of the $b$-hadron, the value of  $P^t_s$ for $b$-quark jets
is higher, as can be seen from fig.~\ref{bt-tv}(c).

{\it The rapidity of each track included in the 
secondary vertex}, $R^{tr}_s$,
is quite a strong discriminating variable, significantly improving
the $b$-quark selection.
Although a $b$-hadron on average is produced with a higher energy, the 
rapidities of particles from $B$-decays are less than those from 
$D$-meson decay, as can be seen from fig.~\ref{bt-tv}(d). This is 
mainly explained by the higher $b$-hadron mass. 
The variable $R^{tr}_s$ is defined for each track in the SV
and the corresponding variable $y_R$ for each track
is used  in (\ref{eq-comb}) for the computation of the $b$-tag.
Although there is overlap between the signal and background for an individual 
track rapidity, because of the  large number of secondary tracks 
the inclusion of all the rapidities in the $b$-tag results in a significant 
gain.

{\it The transverse momentum of an identified energetic lepton}, 
$p^t_l$.
It is independent of the  track IP and can be defined for any
category of jet containing a muon or electron. 
A more detailed description of this variable and of some 
specific features of its inclusion in the combined $b$-tagging scheme are given 
in the next section.

All these variables are defined for the first category of jets (with 
reconstructed SV). For the two other categories, a reduced 
set of variables 
is used. For jets with at least two offsets, the jet lifetime probability,
the effective mass of all tracks with offsets, their rapidities  
and any lepton transverse momentum are computed. For jets with less than 
two offsets the effective mass is not used, as there is no reliable criterion 
for selecting the particles from $B$-decay; however the track
rapidities for all tracks with positive IP are still included
in the tagging. The ratios of probability density functions are computed
separately for each jet category. The possibility of treating 
in the same way different categories of events with different
sets of discriminating variables is a very important feature of the 
likelihood ratio method of $b$-tagging.

The distributions of $M_s$, $X_s^{ch}$, $P^t_s$ and $R^{tr}_s$
are shown in figure~\ref{bt-tv}. These distributions are shown for
$b$-quark jets and also for $c$-quark jets, the latter constituting the main
background for $b$-tagging.

\begin{figure}
  \epsfig{figure=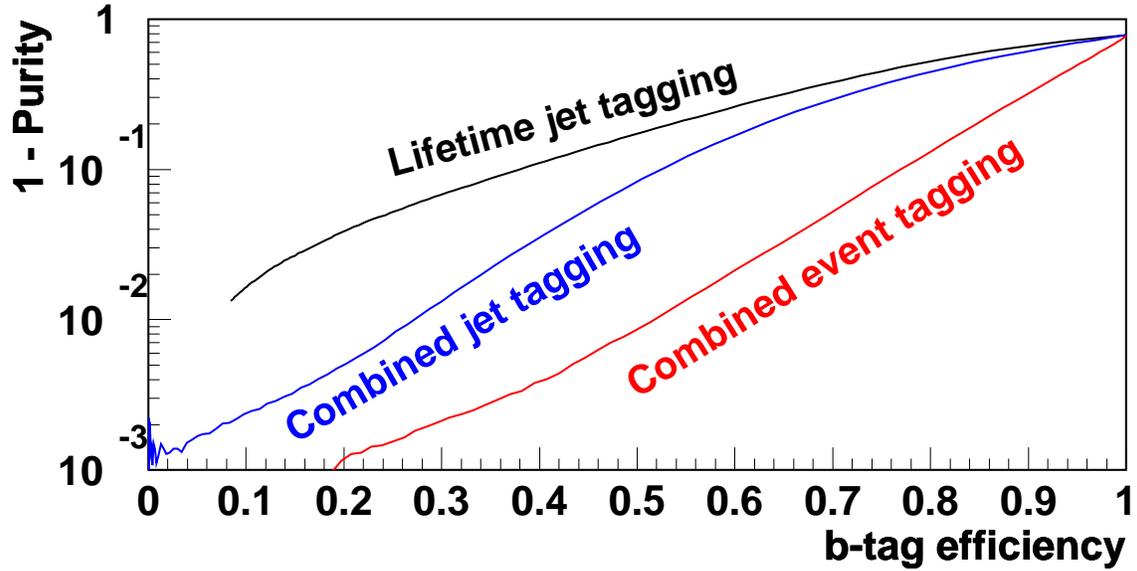,width=17cm}
  \caption{Background suppression in $Z$ hadronic events using
    combined $b$-tagging.}
  \label{bt-cmt}
\end{figure}

Combined $b$-tagging using the complete set of discriminating
variables performs much better than the simple lifetime tagging.
This is illustrated in fig.~\ref{bt-cmt}. The performance is tested using 
jets of $Z$ hadronic decays. The figure shows the contamination of the 
selected sample by other flavours ($N_{udsc}/(N_{udsc}+N_{b}$))
versus the efficiency of $b$-jet selection. 
%
Compared with the tagging using only $P_j^+$, combined
tagging provides much better suppression of background, especially
in the region of high purity. A very pure sample with  contamination
below 0.5\% can be obtained for a sizable $b$ efficiency, which
opens new possibilities for measurements with $b$-hadrons.

\subsection{Lepton Tagging}

Leptons with high transverse momentum have long been used in a variety 
of ways to identify the quark flavour of the jet from which they originate.
This section describes the inclusion of this information 
within the standard DELPHI $b$-tagging algorithm.

Knowing the probabilities $P^{q,c,b}$ of finding a lepton in a light quark-jet,
a $c$-jet or a $b$-jet respectively, and the transverse momentum distributions 
$f^{q,c,b}(p^t_l)$ of these leptons, the contribution of 
identified leptons to the global discriminating variable~(\ref{eq-comb}) is:

\begin{eqnarray}
y^{(c,q)}_{p^t_l} = \left\{ \begin{array}{ll}
  \frac{f^{(c,q)}(p^t_l)}{f^{b}(p^t_l)} 
\frac{P^{(c,q)}}{P^{b}} & \mbox{if a 
    lepton is found,}\\
                     \\
  \frac{1 - P^{(c,q)}}{1 - P^{b}} & \mbox{otherwise,}
                        \end{array} \right.
\label{lepton}
\end{eqnarray}
The quantities appearing in this expression are extracted from 
a sample of simulated 
hadronic $Z$ decays, where all reconstructed particles are clustered 
into jets.

Reconstructed particles are identified as leptons if they 
satisfy a tight electron tag (from energy loss by ionisation in 
the TPC, or from associated energy deposits in the electro-magnetic 
calorimeters), a tight muon tag (from the muon chambers
only), or a standard muon tag confirmed by a minimum 
ionisation energy deposit in the
hadron calorimeters. Detailed descriptions of these different categories of
tags and of their 
performance are given in ref. \cite{delphi}. The quantities $P^{q,c,b}$ are 
simply defined as the fractions of 
jets of the corresponding flavour inside which a lepton is identified, and are
3\%, 9.8\% and 18.7\% for light quarks, $c$-quarks and $b$-quarks respectively.

The transverse momentum of the lepton is evaluated with respect 
to the jet to which it
belongs, when the lepton momentum is subtracted from the 
total jet momentum. The tagging
contribution $y^{(c,q)}_{p^t_l}$ of the leptons is then obtained from 
the ratio of the transverse momentum distributions 
$f^{(c,q)}(p^t_l)$ and $f^{b}(p^t_l)$.

\begin{figure}[htb]
 \begin{center}
   \epsfig{figure=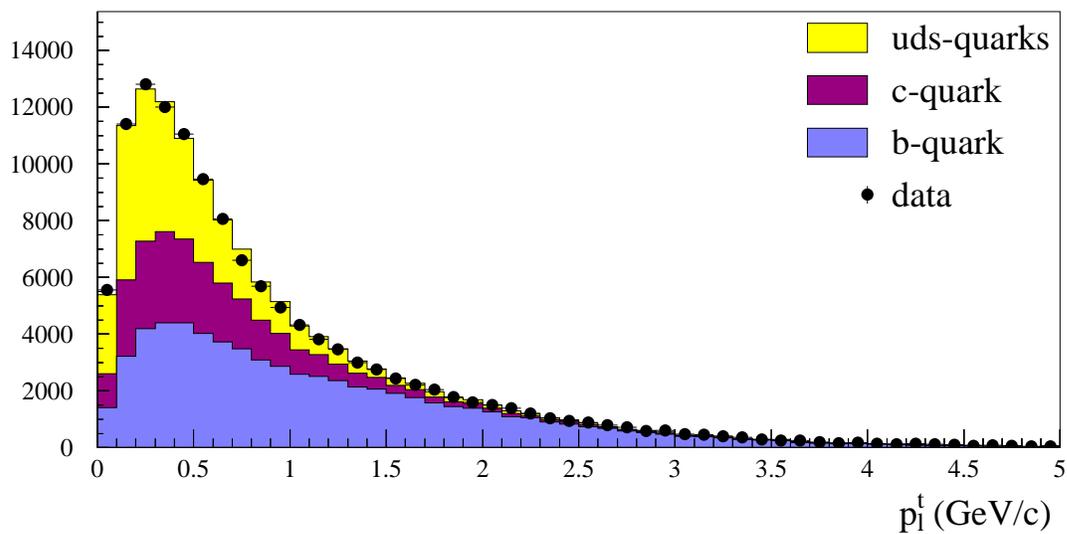,width=16cm}
 \end{center}
 \caption{Transverse momentum distribution of identified leptons with 
	respect to the jet from which they originate, measured in hadronic 
	$Z$ decays at LEP1. The contributions from light 
	quarks, $c$-quarks and $b$-quarks are added and compared to the data.}
 \label{lep_lep1}
\end{figure}

Figure \ref{lep_lep1} shows the superimposed distributions 
of transverse momenta for leptons found in $b$-jets, $c$-jets, and 
light quark-jets. The agreement of these with the data obtained 
from real hadronic $Z$ decays (recorded in 1994) is excellent. 
Because of the small branching ratio of $b$-hadrons to high $p^t_l$ 
leptons, the intrinsic discriminating power 
of lepton tagging is weaker than that of lifetime and secondary vertex 
information. However, it does provide useful information since 
it is fully independent 
of the above, and supplements the $b$-tagging in particular 
when no other hints of $B$ decays are found (for example in case of 
fast decay, or decay outside the vertex-detector acceptance).

\begin{figure}
  \epsfig{figure=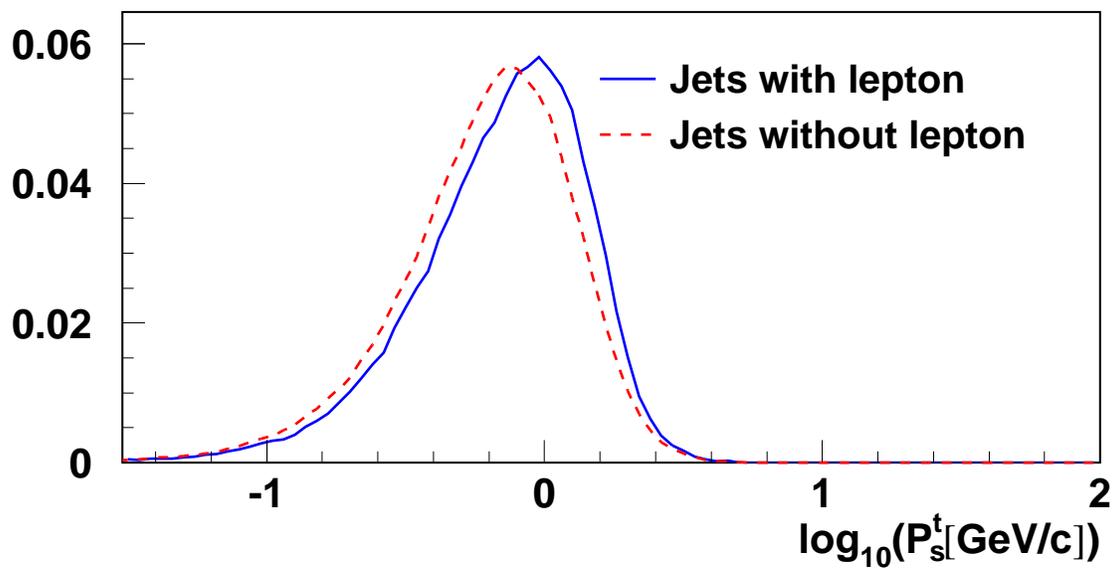,width=17cm}
  \caption{
    Distribution of $\log_{10}(P^t_s)$ for jets with (solid line) and
    without (dashed line) leptons.}
  \label{bt-lept}
\end{figure}

In principle, the distribution in transverse momentum at the secondary vertex
$P^t_s$ depends on whether or not there is a lepton present, because it is
accompanied by an unseen neutrino. However, because semi-leptonic decays are
multi-body, there is not a strong correlation between the lepton and the
neutrino transverse momenta, and the difference in the
$P^t_s$ spectra (i.e. with and without charged leptons) is small, as
can be seen from fig.~\ref{bt-lept}.

The use of the algorithm at LEP2 requires some additional care. 
In this case, data available for detector calibration are less abundant, 
and some experimental aspects are less well understood.
In particular, lepton identification or misidentification probabilities 
are not perfectly reproduced, and this leads in general to an excess 
of simulated events with respect to real data, mainly in the low 
transverse momentum region. To correct for this, rejection factors 
are computed for each category of identified leptons in the simulation, and 
applied randomly. 

\begin{figure}[htb]
 \begin{center}
   \epsfig{figure=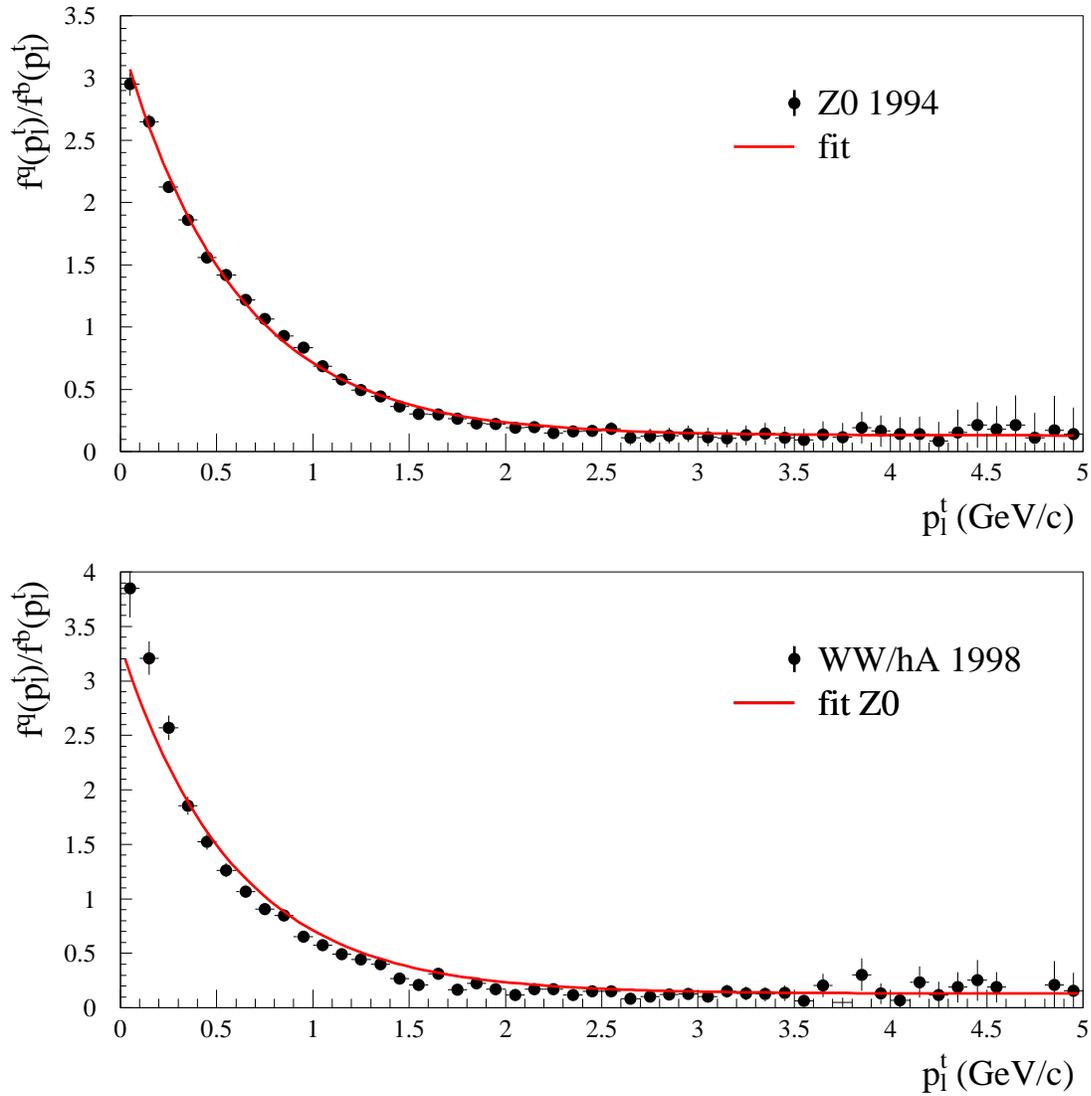,width=16cm}
 \end{center}
 \caption{The upper figure shows the ratio of the numbers of light-quark and 
          $b$-quark events as a function of the lepton transverse momentum 
          (dots), as extracted from simulated hadronic $Z$ 
          decays. The line is the fit to that ratio. The lower figure shows the
          same ratio, extracted from a simulated Higgs signal and WW background
          at LEP2. The line is the $Z$ fit.}
 \label{lr_lep2}
\end{figure}

Finally, it is not obvious that  lepton tagging calibrated on $Z$ data 
will produce efficient discrimination in the context of Higgs 
searches at LEP2, especially in 4-jet events. 
This is because in the Higgs search 
events are reconstructed by constraining the number of jets 
rather than fixing the jet algorithm resolution parameter as was done 
for the calibration; as a result the measured transverse momentum 
distributions are different. Figure \ref{lr_lep2} shows
the distributions of the flavour ratios ($i.e.$ the discriminating power) after
the applied corrections. The data at the $Z$ and at high energy are in fact
in approximate agreement.

\subsection{Event $b$-tagging}

$b$-hadrons are almost always produced in pairs and the presence of the
second $b$-hadron significantly improves the $b$-tagging of an event as
a whole. The likelihood ratio method provides a simple way for combining 
the information from the two $b$-hadrons.
Keeping in mind that each flavour is produced independently from all
other flavours, one can write the equation for the event tagging variable, 
where the two jets\footnote{In events with more than 2 jets, the
smallest two values of $y$ were used.} are of categories $\alpha$ and 
$\beta$, as:

\begin{equation}
y^{ev}_{\alpha \beta} = \frac{R_b}{R_c} 
\frac{n^c_\alpha n^c_\beta}{n^b_\alpha n^b_\beta}
\prod_{i} y_{i,\alpha}^c \prod_{i} y_{i\beta}^c + 
\frac{R_b}{R_q} 
\frac{n^q_\alpha n^q_\beta}{n^b_\alpha n^b_\beta}
\prod_{i} y_{i,\alpha}^q \prod_{i} y_{i\beta}^q
\end{equation}

It was found, however, that a simpler way of combining the information
from two jets into a single tagging variable:
\begin{equation}
\label{eq-effhz}
y^{ev}_{\alpha \beta} = y_\alpha \cdot y_\beta
\end{equation}
works equally well. Here $y_\alpha$, $y_\beta$ are given by eqn.
($\ref{eq-comb}$). The difference between these two equations
is that (\ref{eq-effhz}) neglects the correlated production
of the same background flavours  in an event. The $b$-tagging variable
computed from (\ref{eq-effhz}) was used in the Higgs search (see Section
\ref{Higgs}).

Figure~\ref{bt-cmt} compares the performance of  event tagging and of jet
tagging for $Z$ events. As can be seen, very strong
suppression of background (down to $10^{-3}$ level) can be achieved
with  event tagging.

\subsection{Equalised Tagging}

\label{sect:equalise}

Physics analyses at the edge of detector capabilities, like the search for a
Higgs boson, demand extremely high performance of the $b$-tagging.
The only way to achieve this objective is to expand the set of 
discriminating variables. But adding a new variable in the
combining relation (\ref{eq-comb1}) becomes more and more 
difficult with the growth of their number due to the 
increasing influence of correlations among them. 

\begin{figure}[htb]
 \begin{center}
   \epsfig{figure=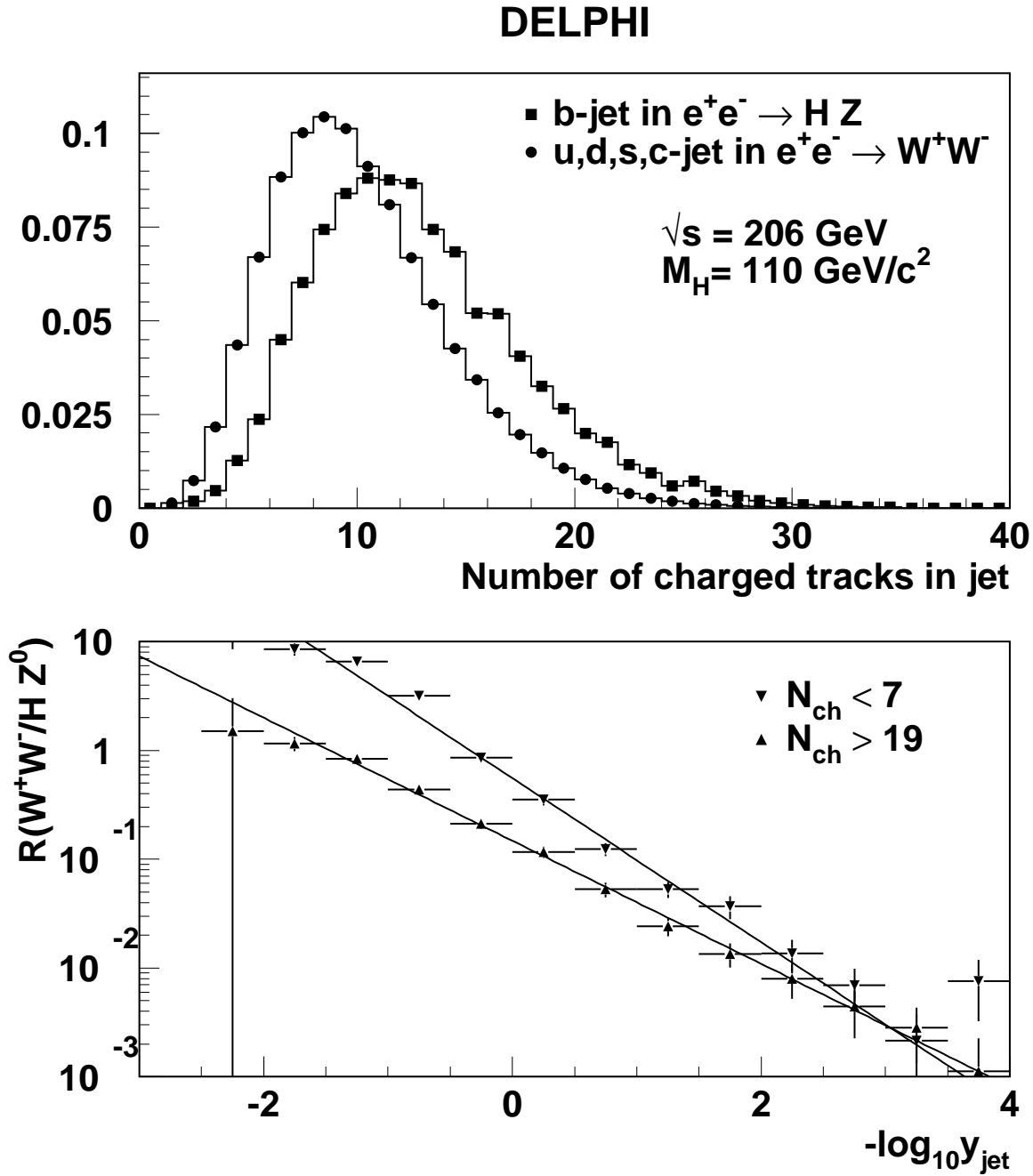,width=17cm}
 \end{center}
 \caption{Upper plot: Simulated distributions of the number of tracks 
   in a $b$-jet from the $e^+e^- \to H Z$ process and in a light 
   quark jet from the $e^+e^- \to W^+W^-$ process. Lower plot:
   the ratio $R(W^+W^- / HZ)$ of the number of light quark jets 
   from the $e^+e^- \to W^+W^-$ process to that of $b$-jets 
   from $e^+e^- \to HZ$ process in the simulation (arbitrary normalisation) 
   as a function of $-\log_{10}y_{jet}$, shown separately 
   for jets with less than 7 or 
   greater than 19 tracks. The lines in each case show the exponential fit 
   of these rates.}
 \label{fig:bt-eq-nt}
\end{figure}

However, the combined method can be modified to include
correlated variables. The main idea of the combined
method which guarantees optimal tagging for non-correlated variables 
consists in assigning the same value of the tagging variable to 
different events having the same likelihood ratio for background to signal. 
As described below, the consistent application of this principle 
while extending the set of discriminating variables gives a desirable 
improvement of the tagging performance. The simplest way to explain
this approach is to consider a particular example, obtained from
simulation samples and presented in figure~\ref{fig:bt-eq-nt}.

The upper plot in fig. \ref{fig:bt-eq-nt} shows 
the simulated distributions of charged multiplicity $N_{ch}$ 
in  $b$-jets from the process $e^+e^- \to H Z$  and  
in  light quark jets from $e^+e^- \to W^+W^-$. 
The latter process presents the main background for the SM Higgs search, 
and is suppressed mainly by the $b$-tagging. The fact that
these two distributions are different implies that it is useful
to include this variable 
in the tagging. The lower plot in fig. \ref{fig:bt-eq-nt} shows the 
ratio $R(W^+W^-/ H Z)$ of the number of light quark jets from 
the $e^+e^- \to W^+W^-$ process to that of $b$-jets from 
$e^+e^- \to H Z$ in the simulation, as a function of the tagging 
variable $X_{jet}$:

\begin{equation}
X_{jet} = -\log_{10}y_{jet}, 
\end{equation}
where $y_{jet}$ is defined by equations (\ref{eq-comb1}-\ref{eq-comb2}). 
Two subsamples of events with $N_{ch}<7$ and $N_{ch}>19$ are considered 
separately. As can be clearly seen, for the same value of $X_{jet}$ the 
ratio $R(W^+W^- / H Z)$ in the two subsamples is different. 
Events with the same value $X_{jet}$ are thus not equivalent; in one 
subsample they will contain more background contamination than in the other.
To restore their equivalence, the variable $X_{jet}$ should 
be modified in such a way that all events with the same value of $X_{jet}'$ 
in different subsamples will have the same ratio $R(W^+W^- / H Z)$. Due 
to the equivalence principle formulated above, such a modification should give
better tagging. Technically, {\it equalising} of $X_{jet}$ is achieved 
by a linear transformation:
\begin{equation}
\label{eq-lt}
X_{jet}' = A\cdot X_{jet}+B, 
\end{equation}
assuming that the dependence of $R(W^+W^- / H Z)$ on $X_{jet}$ in each 
case can be approximated by an exponential, as  shown
in fig. \ref{fig:bt-eq-nt}. The coefficients $A$ and $B$ are different 
for each subsample; their calculation using the parameters of the exponential
functions is straightforward. 

Including a new independent variable $x_{new}$ in the
tagging using (\ref{eq-comb1}) is equivalent to the transformation
$X_{jet}' = X_{jet} - \log_{10}y_{new}$, where 
$y_{new} = f^{bgd}(x_{new})/f^{sig}(x_{new})$, i.e. it is a particular case 
of (\ref{eq-lt}). Such a simple transformation cannot be used for 
$N_{ch}$ because of its strong correlation with other discriminating 
variables, which is reflected in the significantly different slopes 
of the lines in fig. \ref{fig:bt-eq-nt}. Instead, the transformation 
(\ref{eq-lt}) works reasonably well. 

A practical application of the equalising method is to the Standard Model Higgs boson 
search. A set of additional discriminating variables is defined for each jet
of the event. For each new variable, jets are classified in 3 to 5 subsamples.
For example,  for $N_{ch}$ these subsamples are: $N_{ch}<7$;
$7\leq N_{ch}<12$; $12 \leq N_{ch} < 20$; $N_{ch}\geq 20$.
For each subsample the transformation (\ref{eq-lt}) is applied
independently and the new tagging variable $X_{jet}'$ is computed. 
The parameters of the transformation are determined
from the condition that the dependence of $R(W^+W^- / H Z)$ on
the modified $X_{jet}'$ becomes the same for all subsamples.
The variables are included in the tagging sequentially. 
For each new variable, the $X_{jet}'$
obtained at the previous step is used. As before, the global event $b$-tagging
variable $X_{evt}$ is defined as the sum of the two highest $X_{jet}'$ values
among all jets in the event.

The additional variables included in the $b$-tagging using
this equalising method reflect mainly kinematic properties
of $b$-quarks. They are: the polar angle of the jet direction;
the jet energy and invariant mass; the charged multiplicity of the jet; 
the angle to the nearest jet direction; and the number of particles
with negative IP.

\begin{figure}[htb]
  \begin{center}
    \epsfig{figure=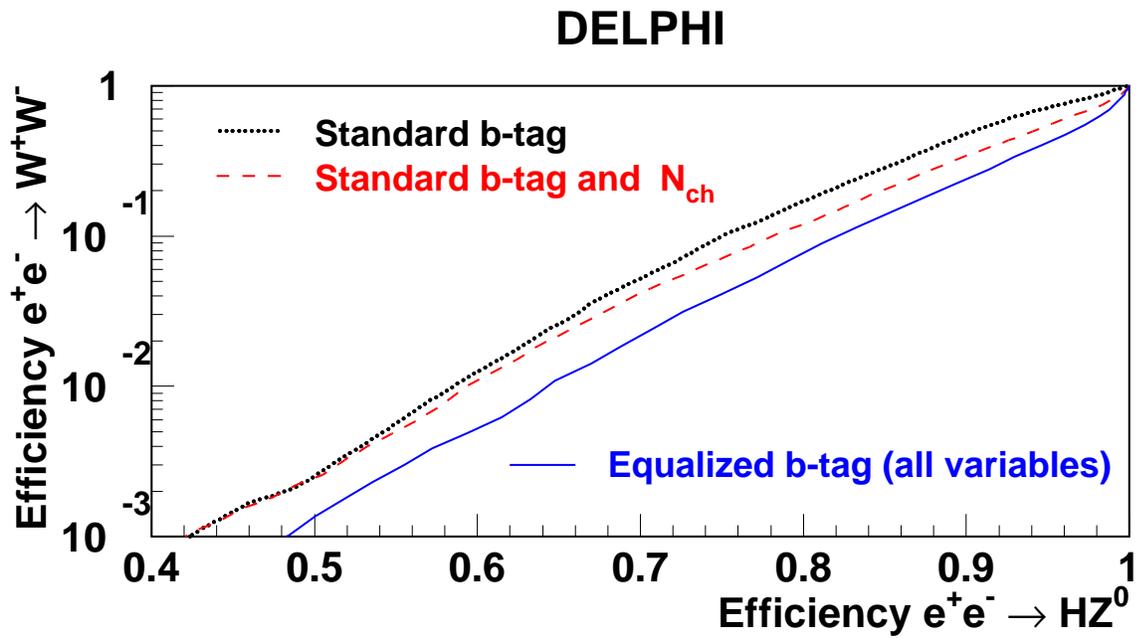,width=17cm}
    \caption{Mis-tagging efficiency for $e^+e^- \to W^+W^-$ versus
      $e^+e^- \to H Z$ selection efficiency, as obtained from
      standard combined $b$-tagging, $b$-tagging equalised with respect to
      $N_{ch}$, and equalised $b$-tagging with the complete set
      of variables (see text for details).}
    \label{fig:bt-eq-perf}
  \end{center}
\end{figure}

Returning to the example of $N_{ch}$, fig. \ref{fig:bt-eq-perf} shows 
that equalising over this variable improves the suppression of the 
$e^+e^- \to W^+W^-$ background. As can be seen 
from fig.~\ref{fig:bt-eq-nt}, the largest difference between 
subsamples with different $N_{ch}$ is observed at low $X_{jet}$ values, 
while the background suppression at high $X_{jet}$ is almost the same. 
The main improvement from the equalising procedure can thus be 
expected for the low purity / high efficiency tagging, corresponding 
to low $X_{jet}$ values. Exactly such behaviour is observed in fig. 
\ref{fig:bt-eq-perf}: including $N_{ch}$ gives almost no improvement for the 
region of strong background suppression. However, equalising the $b$-tagging
for the complete set of variables given above suppresses 
the $e^+e^- \to W^+W^-$ background by an extra factor of more than 2 over a
wide range of $e^+e^- \to H Z$ efficiency.
This additional
suppression is important for the Higgs boson search since it results in
a sizable increase of its detection sensitivity.

The same equalisation procedure was applied for the $e^+e^- \to hA$ 
channel when both Higgs bosons decay into $b\bar{b}$, which is the 
dominant channel with BR larger  than $90\%$ at LEP2 energies.
A new $X_{jet}^{hA}$ was constructed with the condition that 
all events with the same value of $X_{jet}^{hA}$ 
in different subsamples will have the same ratio $R(W^+W^- / hA)$.
The $X_{jet}^{hA}$ variable  was used 
in the search of the hA 4b channel as described in section~\ref{Higgs}.

\subsection{Different Ways of Combining Variables}

As can thus be seen, for combining the separate variables that are relevant
for $b$-tagging, three different methods have been used :

\begin{itemize}
\item The IPs of the different tracks are combined by constructing the lifetime
probability $P^+$ from the probabilities for the significance values of the
various tracks (see Section \ref{sec15});
\item For the different variables contributing to the `combined tag', a
likelihood ratio method is used, as described in Section \ref{descrip};
\item For extra variables, the `equalisation' method of Section
\ref{sect:equalise} is used.
\end{itemize}

To some extent these differences are a result of the historical evolution of
our $b$-tagging algorithms, but there is some underlying logic to these
differences. Thus the likelihood ratio method is guaranteed to give the optimal
signal/background ratio even for correlated variables (assuming of course that the
simulation accurately describes the data, including the correlations). 
However, the method is much simpler when the variables are uncorrelated, and this is how 
the likelihood ratio method was used in `combined tagging'.

This could have been used for
combining the individual IPs, since the error correlations between tracks are 
small. However, it would have been necessary to produce signal/background
probability ratios separately for each class of track (i.e. for each pattern of hits in the
VD). We preferred to use the lifetime probability, where instead the tuning was
performed separately for the different track classes. 

Finally, 
for the Higgs search, extra variables were included to improve the $b$-tagging
performance. Some of these had significant correlations with those already used, 
so they could not simply be added as extra variables in an
extended combined tagging approach. This led instead to the `equalised tagging'. 


\section{Modelling and Tuning of Mass-related Parameters}

\label{mass}


The agreement between data and simulation is sensitive to the modelling
of the physics in the event generator, and to the tuning of its
parameters. A detailed description of the physics model in the main
generator used by DELPHI at LEP can be found in \cite{PB:JETSET}. 
The strategy adopted for the parameter tuning and its corresponding results 
can be found in \cite{QCD-DELPHI}. In this section, two aspects
of the modelling are mentioned, which are specially relevant in the
context of $b$-tagging, and which have been investigated recently 
\cite{PB:YellowReport,PB:Norrbin}: 
the modelling of the rate of gluons radiated off $b$-quarks 
relative to light quarks, and the probability of secondary $b$-quark
production through gluon splitting. Both quantities affect the 
description of the dependence of $R_b$ on the jet multiplicity (and on 
event shape variables), and depend critically on the way quark 
mass parameters are introduced in the treatment of the quark 
fragmentation and subsequent hadronization used in the generator. They 
are important for several of the measurements in Section \ref{applications}, 
particularly those involving the analysis of multi-jet $b$-tagged events, 
such as the measurement performed at LEP1 of the running $b$ quark mass 
at the $M_Z$ energy scale (see Section \ref{sect:mb}), and the LEP2 Higgs 
boson search (see Section \ref{Higgs}).


\subsection{Treatment of Gluon Radiation off $b$-quarks}

\label{gluons}

\begin{figure}[tbh]
  \begin{center}
    \mbox{\psfig{figure=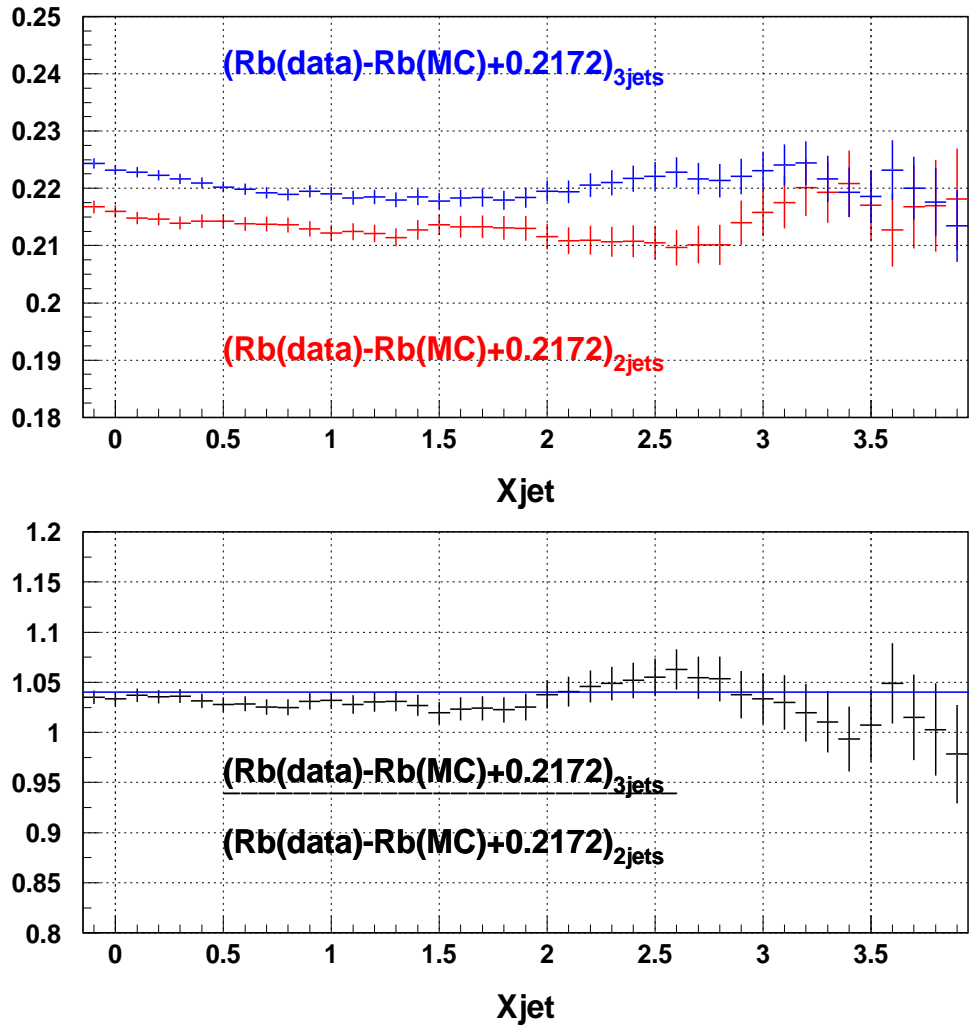, height=15cm, width=15cm,
    bbllx=0, bblly=20, bburx=567, bbury=567}}    
  \caption{Comparison of the measurements of the $R_b$ ratio in 2 and 3
           jet events at the $Z$, as a function of the cut in the $b$-tagging variable
           $X_{jet}$. The simulation used {\tt JETSET 7.4}. 
           \label{PB:rb23}}
  \end{center}
\end{figure}

Discrepancies between simulation and data 
were observed, which could be attributed, entirely or at 
least partly, to imperfect modelling of mass effects in the generator. As an
example, 
the $R_b$ fraction  evaluated separately for two and three jet events, 
using the method described in Section \ref{sect:Rb}, 
is illustrated in figs. \ref{PB:rb23}, 
where {\tt JETSET} version 7.4 \cite{PB:JETSET} was used for the simulation.
Similar behaviour is observed in the comparison of two and four jet events.
                                              
Because of their higher mass, $b$-quarks radiate fewer gluons than
lighter flavours. This results in fewer multi-jet events in the 
case of $b$-quarks. From kinematic arguments, the suppression 
scales approximately as $m_b^2/(s \cdot y)$, where $m_b$, $s$ and 
$y$ are the $b$ mass, the square of the collision energy 
and the jet resolution parameter, respectively \cite{PB:Rodrigo:1999qg}. 
It is observed explicitly in the value of $R_{3}^{bq}$, the double 
ratio of the 3-jet rate for $b$ and light 
quarks,\footnote{ $R_{3}^{bq}$ is defined as the ratio of the $b$-quark and 
light quark rates in 3-jet events, divided by the corresponding flavour 
ratio  for events with any number of jets}
used in the measurement of the running $b$-quark mass at the $M_Z$ energy 
scale (see Section \ref{sect:mb}).
Quantitatively the suppression is of  order  5\%.

In the original {\tt JETSET} prescription, used up to version 7.3, mass effects 
were ignored altogether, both in the parton shower evolution describing the 
fragmentation of the quarks, and in the 3-parton matrix element 
used to correct the first emissions of quarks and anti-quarks 
in the shower. The phase space treatment did include masses, however,  and
induced a large suppression of radiation from the $b$ quark. In version 7.4, 
and later in {\tt PYTHIA} versions up to 
6.130, an intermediate ``improvement" was introduced, in that 
matrix element expressions incorporating quark masses were now used in the 
matching procedure. The suppression of the radiation resulting from this
intermediate treatment, 
which was in place during much of the LEP period, was however exaggerated
by as much as a factor of 2, and
resulted in the largest 
discrepancy with the data \cite{PB:YellowReport,PB:Norrbin}. 
 Starting with {\tt PYTHIA} version 6.130, 
and up to version 6.152, mass effects were also introduced in
the shower evolution through a correction to the expressions 
of the probabilities of the first branchings of each quark. 
From {\tt PYTHIA} version 6.153, a fully consistent treatment is 
available, including a massive treatment of all branchings 
in the shower, now taking into account
in the specification of the matrix element the 
nature of the couplings of the source (vector, 
axial,...) decaying into quarks, as well as the possibility of 
unequal quark masses (as 
in the case of $W \rightarrow c {\bar s}$). Considerable overall improvement
was achieved in the description of both $b$-tagged 3-- and 4--jet  rates,
thanks to these developments \cite{PB:YellowReport,PB:Norrbin}.


\subsection{Treatment of Gluon Splitting to $b$-quark Pairs}

Secondary $b$-quark pair production from gluon splitting can also result 
in $b$-tagged multi-jet events. The corresponding rate is small but is
poorly known both theoretically and experimentally. This implies an 
uncertainty in the predictions, particularly of the $b$-tagged 4-jet rate 
at LEP2 energies.

Measurements by LEP and SLD collaborations at $\sqrt{s} = M_Z$ give 
$g_{b\bar{b}} = (0.254 \pm 0.051)\%$~\cite{lephf}, where
$g_{b\bar{b}}$ is defined as 
the fraction of hadronic events containing a gluon splitting 
to a $b\bar{b}$ pair. This is consistent with 
the best theoretical estimates, which are around 
0.2\%~\cite{PB:YellowReport}, with relative uncertainties 
due to unknown sub-leading logarithmic corrections, which may be as
large as 30\%. 

The rate predicted by Monte Carlo generators 
based on parton shower methods is also sensitive to the treatment of 
sub-leading and kinematic effects in the shower evolution. While the 
original {\tt JETSET} and {\tt PYTHIA} prescription resulted in only 
0.15\%, since version 6.131 a set of new options has been introduced
which bring this rate closer to the measured values~\cite{PB:YellowReport}. 
Two of these options, which almost exactly double the original rate, 
have been recommended~\cite{PB:YellowReport} and are used in 
the latest simulations at 
LEP2 energies\footnote{For simulations performed with
versions of {\tt PYTHIA} prior to 6.131, a re-weighting procedure
was used to increase the $g \rightarrow b {\bar b}$ rate by a factor of about 
2.}. The first of these options ({\tt MSTJ(44)}=3) 
uses the mass of the virtual gluon involved in a splitting 
to define the scale $m_g^2/4$ relevant for $\alpha_s$, the strong coupling 
constant, rather than the default $p_T^2$ prescription used for other types 
of branchings in the shower evolution. The second option ({\tt MSTJ(42)}=3)
reduces the conditions on coherence in the emissions, in the case of gluon 
splittings into heavy quark pairs, by introducing a mass correction into 
the angular criterion used to restrict the successive branchings in the shower.

The impact on the $b$-tagged 4-jet rate at LEP2 is best illustrated
in the context of the Higgs search or of the measurement of $Z$ boson pair
production. At LEP2 energies the 
rate of gluons which can split into $b$ quark pairs is less suppressed by
kinematics than at LEP1. For instance at 189 GeV it is  as large as 
0.4\% using the original {\tt JETSET} and {\tt PYTHIA} prescription. In
a subsample of events enriched with 4-jet events, it can reach levels near
1\% depending on the criterion used on the jet resolution parameter.
With the new options described above, these values are roughly doubled. 
The effect of this doubling on the Higgs search was studied by 
comparing the numbers of events predicted to be selected, when assuming 
the default value for the rate of gluon splittings into $b$ quark pairs or 
the doubled one. The relative difference between these two numbers 
is about 2.5\%; it varies slightly  with the $b$-tagging cut but
does not exceed 5\%. It was taken into account in the final 
evaluation in this channel (see Section 6.5 and references therein).


\section{Physics Applications}

\label{applications}

In this section, some analyses involving $b$-tagging are described. The aim is
not to present the physics results, which have already been published, but
rather to illustrate how $b$-tagging works in practice. The extent to which 
uncertainties in tagging influence the final results is also mentioned. 
Much more detail, including the estimation of systematics, can be found
in the published papers.

\subsection{Measurement of $R_b$ at the $Z$}

\label{sect:Rb}

One of the most challenging measurements at LEP1 is the determination
of $R_b$, the branching ratio for  $Z$ hadronic decays into $b$-quarks. 
All accurate measurements of $R_b$ 
use the so-called double tag method. This compares the
fractions of events in which there is a $b$-tag in a single hemisphere with
those in which both hemispheres are tagged. It allows the 
extraction from the data of both $R_b$ and the efficiency $\epsilon_b$ for 
tagging a hemisphere as coming from a $b$-quark. 
We thus do not have to rely on the simulation
for the calculation of this important quantity.
The small background mis-tag rates, $\epsilon_c$ and $\epsilon_{uds}$, and the 
correlation between hemispheres, $\rho_b$, are taken from the simulation.
The tracks are separated into two hemispheres by the plane perpendicular to the
thrust axis. The highest $b$-tagging value 
$X_{jet} = -\log_{10} y_{jet}$ (see eqn. (\ref{eq-comb})) 
of any jet in the given hemisphere is taken as the hemisphere tag.

The correlation $\rho_b$ allows for the fact that there are small differences 
between the
overall hemisphere $b$-tagging efficiency, and the efficiency for tagging a
hemisphere if the other one has already been tagged. These correlations arise,
for example, from the fact that at the  $Z$, hadronic events tend to consist 
of back-to-back jets; if one jet is at large positive cos$\theta$ where the 
tagging efficiency is lower (see fig. \ref{chiarathetaplot}), then the
other jet is likely to be at large negative cos$\theta$, again with lower
efficiency; thus the efficiencies are correlated.
The systematic uncertainty on $\rho_b$ was estimated by comparing data and
simulation for various kinematic variables that were sensitive to the
separate contributions to $\rho_b$.

In particular, it was found that a large correlation arose from the use of a
common PV for the whole event; if the PV was badly measured as 
being closer to the SV in one hemisphere, then the IP values 
in that hemisphere would all be systematically reduced in magnitude, 
while those in the opposite hemisphere would be increased. 
This was overcome by determining a separate PV for each
hemisphere of the event. This modification slightly reduced the flavour
discrimination power of the algorithm, and hence increased 
the statistical error, in exchange for a large decrease in 
the correlation  and a smaller systematic error.

Achieving high accuracy for $R_b$ requires the following:
\begin{itemize}
\item 
the $b$-tag must reach very high efficiency to reduce the statistical 
      error: $\delta R_b \sim 1/\epsilon_b$;
\item 
at the same time the $b$-tag must have high purity
to reduce the systematic errors coming from our knowledge 
of the  background:  $\delta R_b \sim \epsilon_x R_x / \epsilon_b$, where
$x = q$ or $c$;
\item  
there must be excellent agreement between data and simulation 
to reduce the systematic errors due to the modelling of the detector 
resolution, and because there are quantities taken from the simulation 
and not measured directly in the data.
\end{itemize}

In the DELPHI $R_b$ measurement, both the 
crucial high-purity $b$-tag in the ``multivariate"  analysis\cite{gbb}, 
which finally gave the
best precision, and the ``combined $b$-tag'' analysis 
used the combined hemisphere tag described earlier (see 
Section \ref{sec3}). This required the presence of a SV and included 
the hemisphere lifetime probability $P_j$, the SV mass $M_s$, the charged 
energy fraction $X_s^{ch}$, and the rapidities of the tracks at the SV. The missing
transverse momentum $P_s^t$ and the lepton transverse momentum $P_l^t$ were not
used.

The sources of the systematic error are our knowledge of $b$-hadron, 
charm- and light-quark physics (such as lifetimes, decay modes and
multiplicities) and our understanding of the detector resolution. 
The first contribution is minimised by measuring the $b$-efficiency 
from the data
itself and by reducing the charm- and light-quark mis-tag rates to the minimum
possible so as to give a very pure $b$-tag. 
The second contribution is  minimised by having good agreement between 
data and simulation for the detector resolution. 

The high statistical precision of the result 
is mainly due to the high performance of the DELPHI $b$-tag: 
at 98.5\% hemisphere $b$-purity, the hemisphere $b$-efficiency is 29.6\%,  
while the mis-tag efficiencies $\epsilon_c=0.4\%$ and $\epsilon_{uds}=0.05\%$.

The smallness of the systematic error comes specifically from 
the fact that the contribution from the detector resolution understanding 
is very small.
\noindent
First, the DELPHI vertex detector has three layers of silicon detectors
allowing better pattern recognition than for a detector with only 
two layers. The design of the
detector is such that the intrinsic IP resolutions in both  the $R\phi$ and  
$Rz$ components are good  -- 27 and 39 $\mu$m respectively -- and consequently
also the precision of the primary and secondary vertex 
positions (see Sections \ref{sec12} and \ref{sec22}). 
\noindent
Secondly, the detector IP resolution  is tuned
with high accuracy, as described in section \ref{tuning}, resulting in
a good agreement between data and simulation (see fig.~\ref{tun:fig3}(b)).
As a consequence the error coming from our understanding of 
the detector resolution amounts to only 20\% of the total error.

\begin{figure}[htb]
\begin{center}
\mbox{\epsfig{file=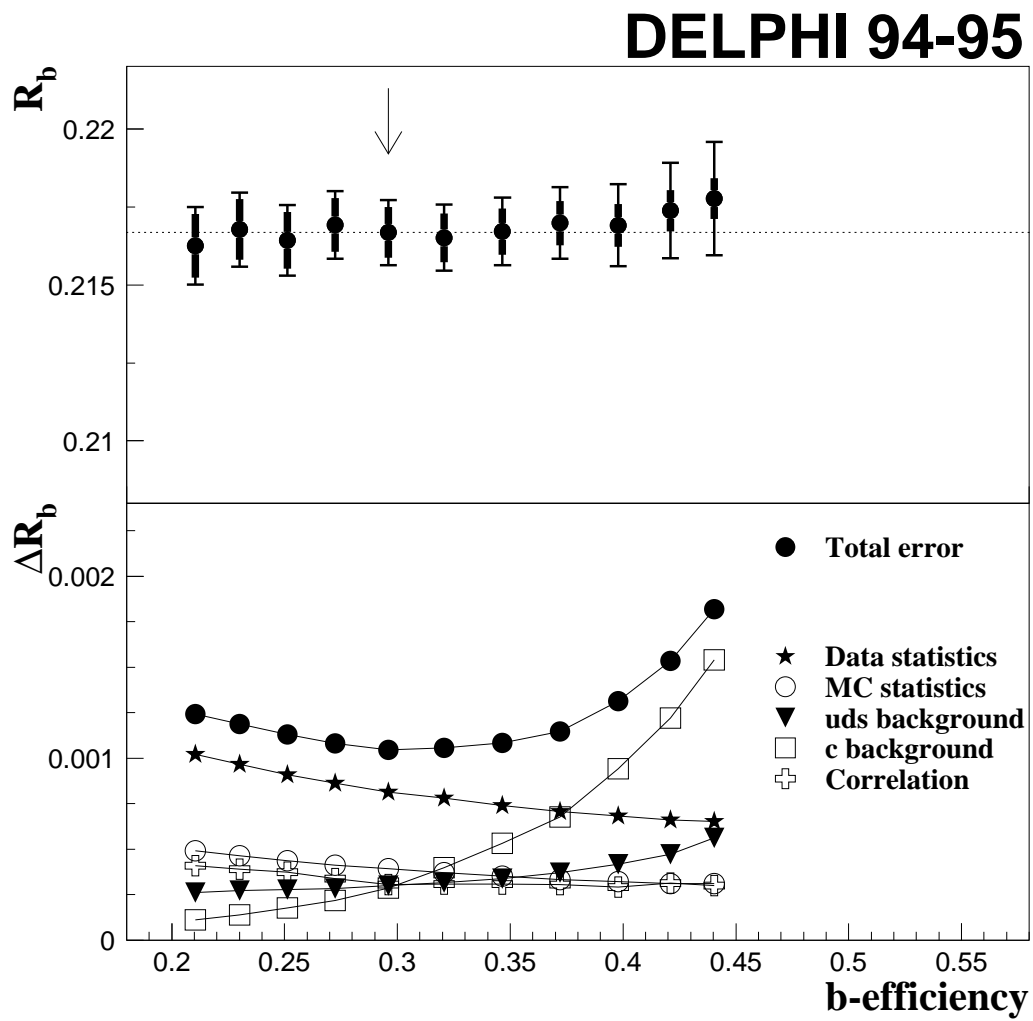,width=15.cm}}
\caption{ Stability of the R$_b$ result as a function of the 
b-tagging efficiency, for data collected in 1994--5. The arrow shows the
$b$-efficiency chosen for the final result.}
\label{rb_stab}
\end{center}
\end{figure}

In summary the performance of the $b$-tag and the understanding of the 
detector resolution result in  very good stability of the $R_b$ measurement
as a function of the $b$-efficiency as shown in fig. \ref{rb_stab}; 
the highest and lowest efficiencies shown of 44.0\% and 21.0\% correspond to
$b$-purities of 91.6\%  and 99.4\% respectively. Thus, a stable $R_b$
result was obtained while the background contribution varied
by more than a factor of 10. The total relative error was only 0.4\%.

\subsection{R$_{4b}$, the Rate of Events with 4 b-quarks at the $Z$}

Four $b$ jets are produced predominantly when, in an event with a $b$ pair, a
gluon is radiated from one of the quarks and itself produces another $b$ pair.
This analysis thus gives information on the $gb\bar{b}$ coupling.

\begin{figure}[htb]
\begin{center}
\mbox{\epsfig{file=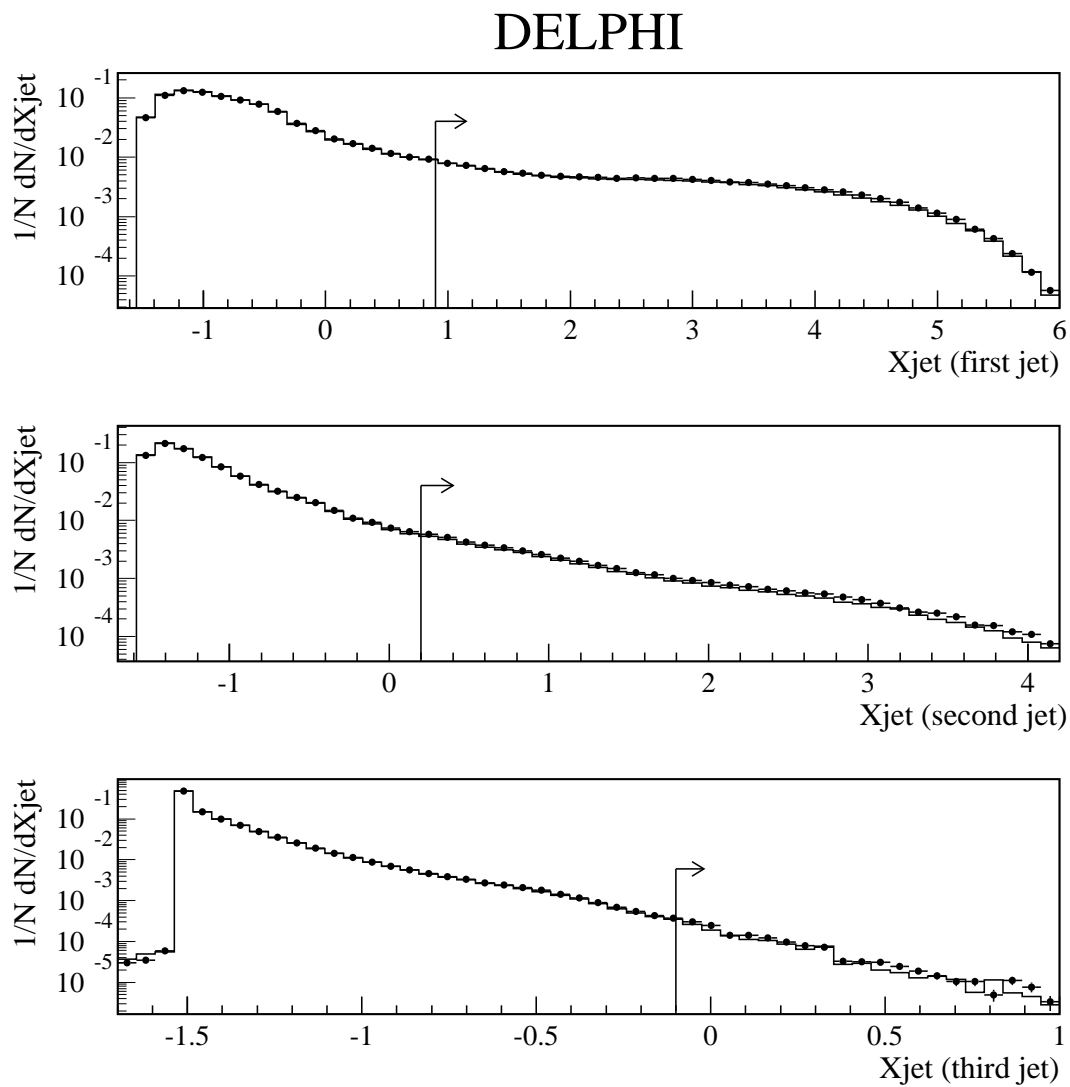,width=14.cm}}
\end{center}
\caption{ Distribution of the $b$-tagging variables for the first three
     jets, ordered according to their $b$-tagging variable, for $Z$ data 
      (dots) and simulation (histogram). The arrows show the positions of the
cuts used to select the $b$ jets. The figure is from reference \cite{r4b}.}
\label{xeffj}
\end{figure}

The high purity and efficiency of the tagging method, together with the good
agreement between data and simulation, allowed DELPHI to measure
for the first time the rate of $Z$ events with 4 $b$-quarks in the 
final state \cite{r4b}. $R_{4b}$ has been measured to be 
(6.0 $\pm$ 1.9 ({\sl stat}) $\pm$ 1.4 ({\sl syst})) $\times$ 10$^{-4}$ 
at a signal efficiency  $\epsilon_{4b}$=(3.16$\pm$ 0.11)\%.
The analysis required 3 jets identified as coming from $b$-quarks. This means 
that the $b$-efficiency enters to the third power, demanding
a very high efficiency of a tight $b$-tag. The high purity of the tag is 
required in order to suppress the background from  gluon splitting into 
$c$-quarks, that is a factor of 10 higher than the splitting into $b$-quarks. 
Thus, the measurement relies on the $b$-tagging performance and 
data/MC agreement in the $b$-tagging of the third jet, 
sorted by decreasing $b$-tag value, see fig. \ref{xeffj}.
The uncertainty coming from  the $b$-tag amounts to only 
6\% of the total systematic error and is determined mainly by the IP resolution
description.

This analysis used the $b$-tag algorithm  of Section \ref{sect:Rb}, except that
if no SV was found, the jet lifetime probability $P_j$ was used by
itself.

\subsection{Measurement of the $b$-hadron Charged Multiplicity}

The good agreement between data and simulation achieved by the tuning 
of the track resolution allowed DELPHI to measure
with very high precision the charged multiplicity $n_B$ of weakly decaying 
$b$-hadrons \cite{nb}.
The basis for the measurement is  the determination of the number
of tracks in a $b$-jet which come from the SV rather than from the PV.  
\noindent
In the hemisphere opposite to the $b$-tagged one, the difference
\begin{equation}
       N_{+-}= n^+ - n^-
\end{equation}
is computed, in which 
$n^+$ and $n^-$ are the numbers of tracks with positive and with negative IP 
respectively. 
The  quantity $N_{+-}$ is obtained as a function of the $b$-tagging purity and 
the value of $n_B$ is extracted by comparing $N_{+-}$ from data and simulation,
extrapolated to  the limit of 100\% $b$-purity.

The result was $n_B$ = 4.97 $\pm$ 0.03 $\pm$ 0.06.
The measurement reaches 1.3\% precision, due to the good 
understanding of the IP resolution, to the efficient method 
for determining the sign of the IP (see Section \ref{sec13})
and to the precision of the VD alignment (see section \ref{align}). 
The tracking efficiency  is $99 \pm$1\%(syst), its uncertainty
dominates the systematic error on $n_B$.

This analysis used the same $b$-tag algorithm as the previous analysis.

\subsection{\bf Measurement of the Running \boldmath $b$-quark Mass at $M_Z$}

\label{sect:mb}

The $b$ quark mass determination at the $M_Z$ scale has been performed by
DELPHI by measuring the $R_3^{bq}$ observable, as defined
in Section \ref{gluons}.
Two different jet-finding algorithms, {\sc Durham} \cite{durham}
and {\sc Cambridge} \cite{camjet}, were used to
reconstruct the jets.    
Special features of this analysis in connection with the flavour tagging 
performance of DELPHI are:

\begin{itemize}

\item[$\bullet$]
$b$ and $q$ initiated events are selected using the same technique,

\item[$\bullet$]
the efficiency versus purity working points are chosen 
so as to minimise the total error on the result (including effects from
corrections and biases).
\end{itemize} 

\begin{table}[htb]
\begin{center}
\vspace{7mm}
\begin{tabular}{|c|c|c|c|c|}
\hline
Method & Tagged Sample & Actually $q$ (\%) & Actually $c$ (\%) & Actually $b$  (\%)\\
\hline
IP & $q$ & 85.8 & 12.6 & 1.7 \\
IP & $b$ & 5.0 & 15.3 & 79.7 \\
\hline
Comb  & $q$ &82.0 &15.5 & 2.5 \\
Comb  & $b$ & 4.3 & 10.4 & 85.4 \\
\hline
\end{tabular}
\vspace{0.5cm}
\caption{Flavour compositions of the samples tagged as $q$-quark 
and $b$-quark events for each tagging method.}
\label{tab:ciq}
\end{center}
\end{table}

The directly measured observable, $R_3^{bq-meas}$, had to be corrected for
detector acceptance effects, for kinematic biases introduced by the tagging
procedure and for the hadronization process in order to get the observable at
the parton level, $R_3^{bq-par}$ \cite{delmbmz}. This quantity was then
compared with theoretical predictions based on `Next to Leading Order'
analytic calculations \cite{PB:Rodrigo:1999qg}  to evaluate the $b$-quark mass at the
$M_Z$ energy scale.

In the original analysis \cite{delmbmz}, $q$- and $b$-quark initiated events
were selected by the lifetime-signed IPs of charged particles 
in the event (see Section \ref{sec15}). In more recent versions of the
analysis, the combined tagging technique of Section \ref{sec3} has also been 
used.
The flavour composition of the samples tagged as $q$-quark and as $b$-quark
by the two methods are shown in Table \ref{tab:ciq}.  

\begin{figure}
  \begin{center}
    \epsfig{figure=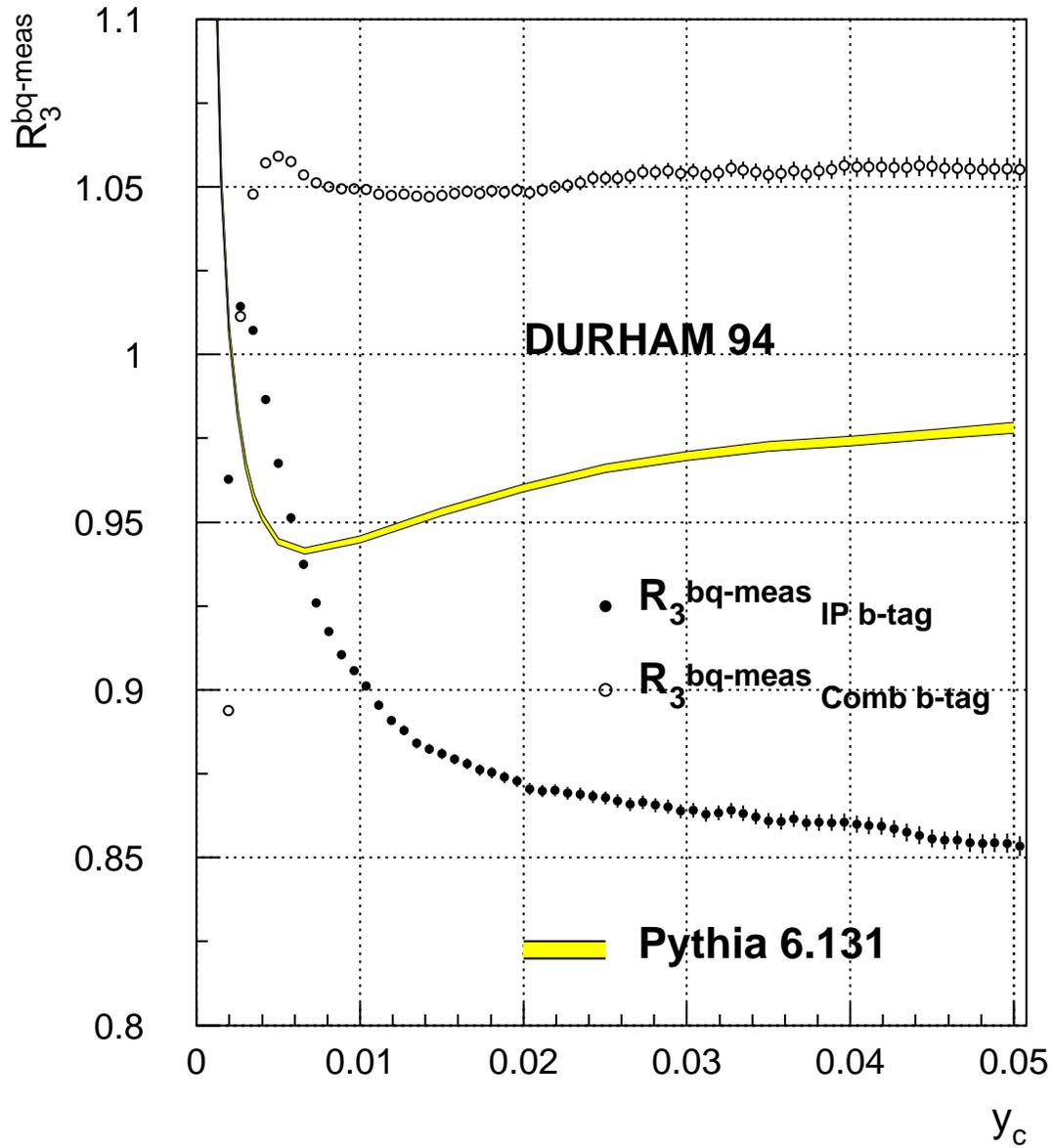,width=14cm}
    \caption[]{The $R_3^{bq-meas}$ observable from simulated DELPHI data 
    for the two tagging techniques and the parton level 
    $R_3^{bq-part}$ obtained with {\tt PYTHIA} 6.131.  
    The {\sc Durham} jet finding algorithm is used.}
    \label{fig:rn_rec_nim_3d_94_v710}  
  \end{center}
\end{figure}

The magnitude of the corrections applied in each of the two $b$-tagging
techniques is illustrated in fig.~\ref{fig:rn_rec_nim_3d_94_v710}, 
where the corresponding $R_3^{bq-meas}$ 
observables are shown, using simulated DELPHI data and the DURHAM jet 
finding algorithm, together with the parton-level one, $R_3^{bq-par}$,
obtained with {\tt PYTHIA} 6.131. For $y_c = 0.02$, the corrections are ~10\% 
for both techniques, although in opposite directions. 

For the simulation the same corrected result is
(obviously) obtained, but for real data this is not the case. This difference
between the two techniques arises mainly from the imperfect modelling of
the physics processes affecting the fragmentation and decay of heavy and
light quarks. 
Half of this difference is taken as the systematic uncertainty 
associated with this measurement.
The error induced by the tagging uncertainties is in the range 0.3\% to 0.4\%,
compared with the total error of order 1\%
on the flavour independence of the strong coupling constant.
In terms of the running $b$-quark mass, these errors correspond to 150 and
500 MeV respectively.

\subsection{Higgs Searches in 4-jet Topology}

\label{Higgs}

One of the main topics  at LEP2 energies has been the search for
the SM Higgs boson, both in the Standard Model (SM) and in the Minimal
Super-Symmetric Model (MSSM).
Here the approach used in the dominant 4-jet channel is outlined.

Radiative production from a virtual $Z$ 
$e^+ e^- \to Z^* \to H Z$ is (in principle!) the main  Higgs
process at LEP2 and is referred to as Higgsstrahlung. 
The mass of the Higgs boson is at present unknown,
but for a given mass its other properties are determined from the SM.
The Higgs boson couples to massive fermions and to the $W^\pm$ and 
$Z$ bosons.

The predominant decay mode for the SM Higgs in the 
mass  range of interest for LEP2 searches  is expected
to be to pairs of $b$ quarks
with a branching ratio ranging from 87\% to 80\% with increasing Higgs  mass.  
The same decay modes are dominant for $h$ and $A$ 
for many choices  of values of the 
parameters in the MSSM, in particular for $\tan \beta > 1$. 
The identification of $b$ jets and rejection of non-$b$ jets is the most
important ingredient in the majority of  analyses designed to search for
the neutral Higgs boson.

The four-jet final state includes the Higgs boson decaying
to $b\bar{b}$, and also in principle to  $q\bar{q}$ or $gg$.
It is characterised by a large amount of
visible energy. As the Higgs boson decays mainly to $b$ quarks,
when the $Z$ also decays to $b\bar{b}$  the event 
will contain 4 $b$ jets. If instead the $Z$ decays to other quark pairs, the
topology will again be 4 jets, but with only two of them due to $b$'s.

The main backgrounds are two-fermion processes
$e^+e^- \to q\bar{q} (\gamma)$,  and four fermion processes involving
$W^+W^-$ and $ZZ$. Pair production of $W^\pm$ can result in
$c$ jets, but only very rarely in $b$ jets. The $ZZ$ is an irreducible
background if  the masses of the Higgs boson and of the $Z$ are close,
and the $Z$ decays to  $b\bar{b}$. The cross-sections of the background
processes are much higher than that for the Higgs 
production. At the highest LEP2 energies (around 208 GeV)
and for $m_H= 114 \;$ GeV/c$^2$, $\sigma_{Hqq}=0.072$ pb while
$\sigma_{4f}=19$ pb and $\sigma_{2f}=78$ pb.

In  searches for the Higgs bosons at LEP2 in DELPHI, the various differences  
between $b$ jets and light quark jets were accumulated in a single variable
$X_{jet}$ defined for each jet, as described
in Section \ref{sec3}. Extra variables which help
discriminate between the signal and background were included in the
construction of the equalised tag (see Section \ref{sect:equalise}).
Including these extra variables in the tagging algorithm significantly improves
the rejection of the light quark background.

\begin{figure}[htb]
  \begin{center}
    \epsfig{figure=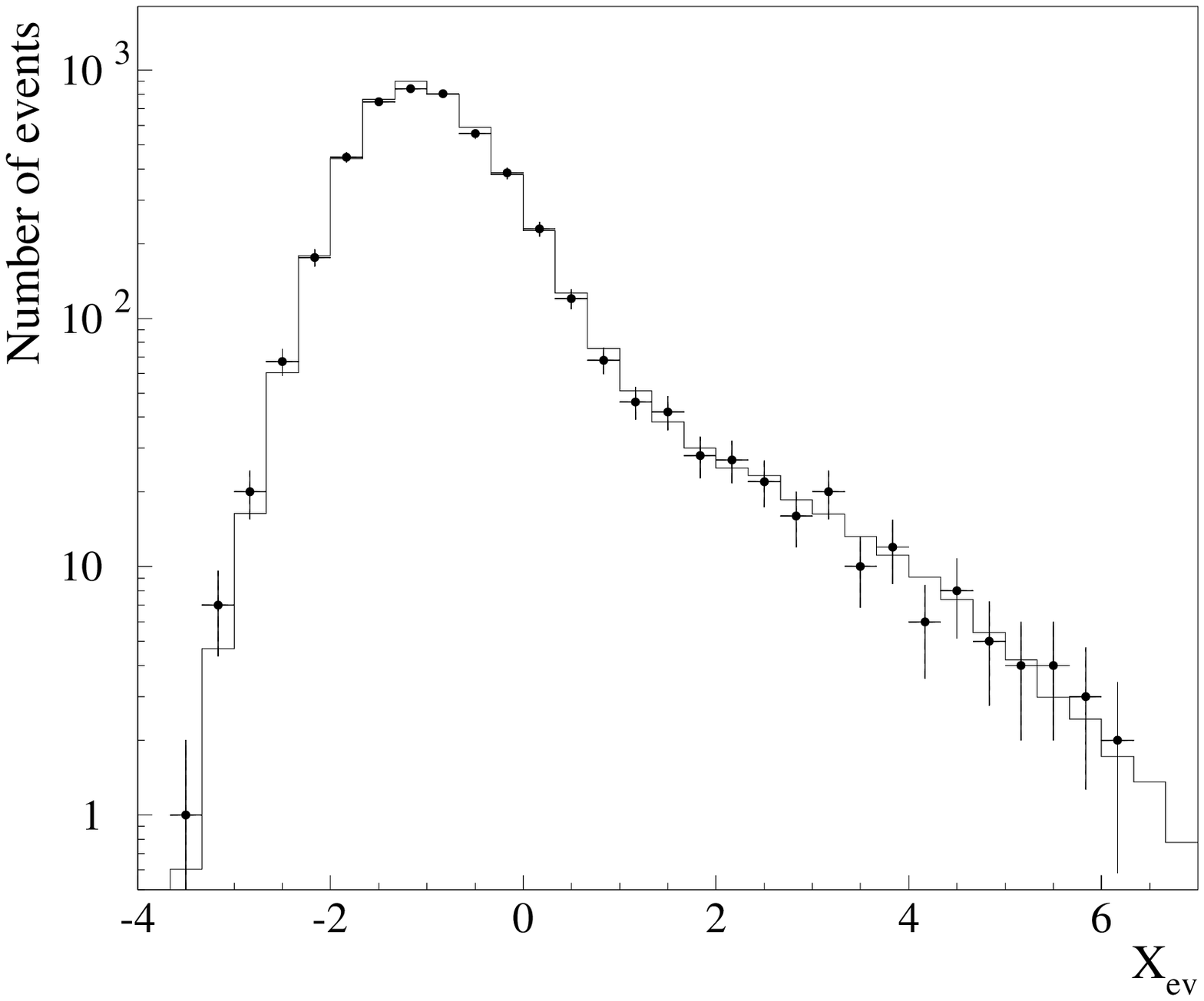,height=10cm,width=17cm}
    \epsfig{figure=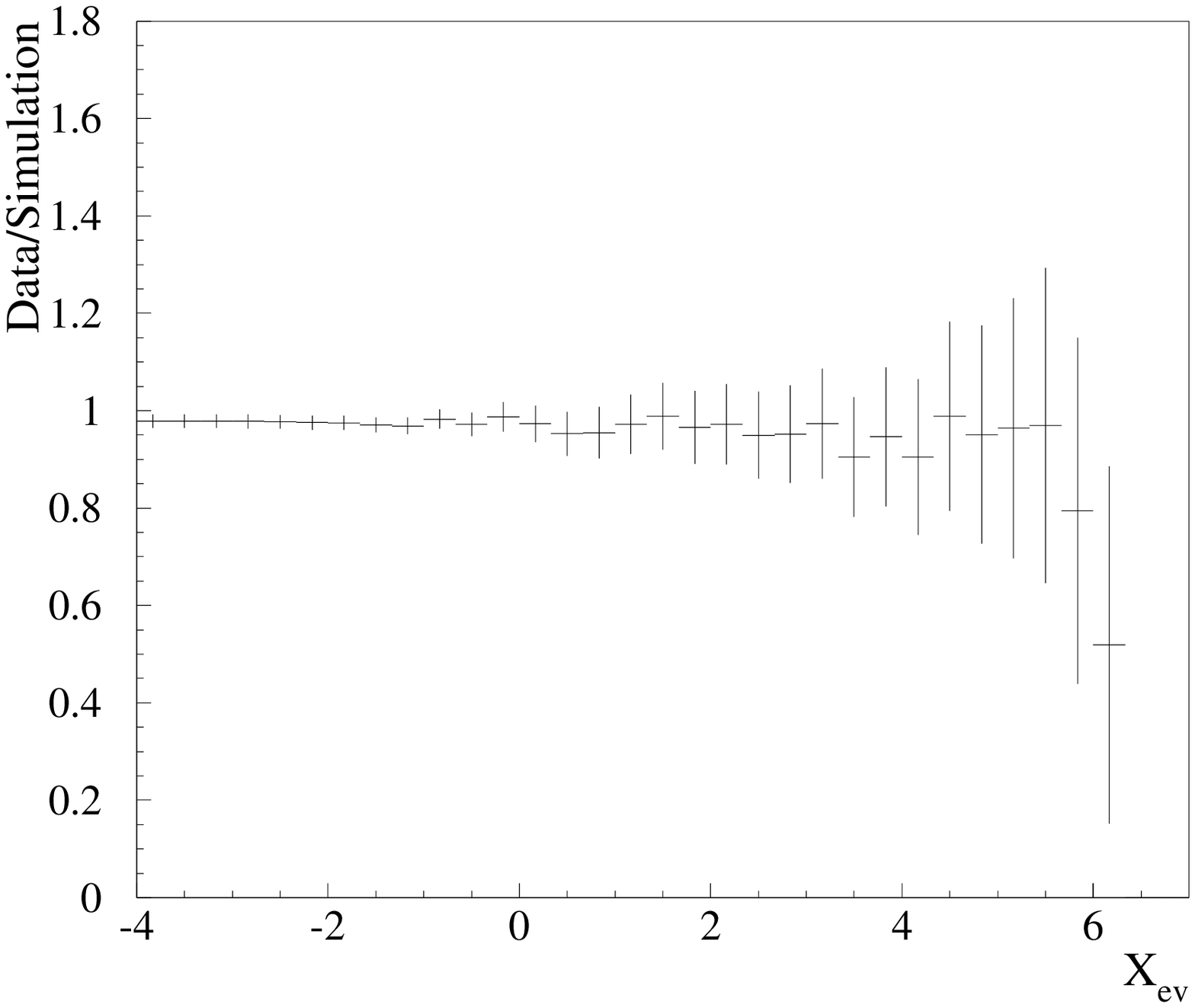,height=10cm,width=17cm}
    \caption[]{Top: distribution of the equalised $b$-tagging variable for
      4-jet events
      at $\sqrt{s} \  = \  192 - 210 \ $ GeV, data (dots) and simulation
      (solid line). Bottom: ratio of the integrated tagging rates in data
      and simulation as a function of the cut in the 
      equalised $b$-tagging variable. The agreement between data and
      simulation is satisfactory.}
    \label{fig:9900}
  \end{center}
\end{figure}

The  $b$-tagging value $X_{ev}$ of the event in the search for 
the SM Higgs boson is defined as
the maximum $b$-tagging value for any di-jet in the event,
computed as the sum of the corresponding jet $b$-tagging values.
In figure~\ref{fig:9900} the distribution of this equalised $b$-tagging
variable is shown after the common four jet preselection \cite{fastpaper}
where good agreement between data and background simulation can be
observed.

\begin{figure}[htb]
\begin{center}
\epsfig{figure=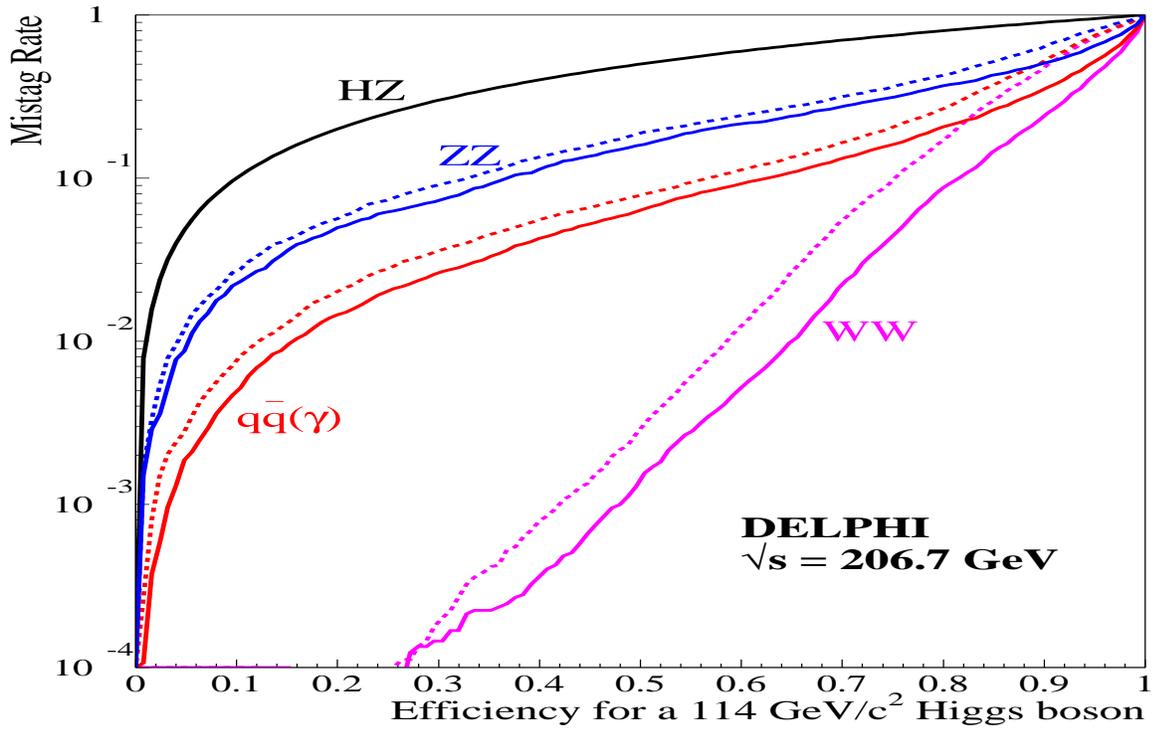,height=10cm,width=17cm}
\epsfig{figure=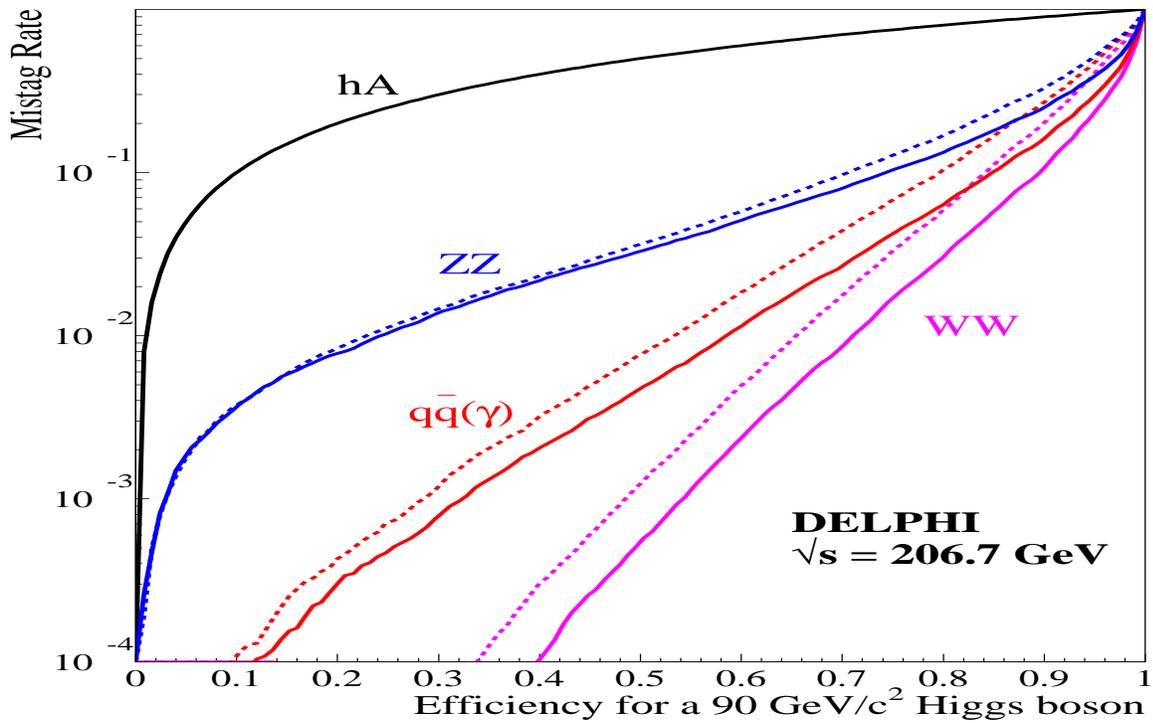,height=10cm,width=17cm}
\caption[]{Expected SM background mis-tag rates 
at $\sqrt{s} \; = \; 206.7 \;$ GeV, as  functions of the signal efficiency 
for a Standard Model Higgs boson of mass 114 GeV/c$^2$ (top) and 
MSSM Higgs boson of mass 90 GeV/c$^2$  and  $\tan \beta=20$ (bottom) 
when varying the cut on the equalised $b$-tagging variable 
(solid lines) and combined $b$-tagging variable (dotted lines). 
}
\label{fig:perfeq}
\end{center}
\end{figure}

In the top part of
figure~\ref{fig:perfeq} the performance of combined and equalised
methods are compared for a SM Higgs boson mass of 114 GeV/c$^2$.
As an example using only the $b$-tagging variable, 
for a signal efficiency of $40\%$ the  $q\bar{q} (\gamma)$ contribution
is reduced to $5.6\%$ using combined $b$-tagging, wheras it is even
more strongly suppressed to  $4.3\%$ 
using equalised $b$-tagging. The $WW$ component is
reduced twice as much using equalised as compared with combined $b$-tagging.


In the bottom part of figure~\ref{fig:perfeq},
the performance of the hA equalised $b$-tagging is shown 
in the $hA \to 4b$ channel and compared with the performance of 
combined $b$-tagging. The presence of four $b$-jets in the signal makes 
the analysis of the $hA \to 4b$ channel different from the  $HZ$ case,
where in most of the cases only two $b$-jets are present. The event 
$b$-tagging value $X_{ev}^{hA}$ for the Higgs boson search in 
the $hA \to 4b$ channel is defined as the minimum  $b$-tagging value 
for any di-jet in the event and is computed as the sum 
of the corresponding jet $b$-tagging values. As can be seen,
the application of the hA equalised b-tagging improves significantly 
the performance of the $hA \to 4b$ channel selection.

As an example using only the $X_{ev}^{hA}$ $b$-tagging variable, 
for a signal efficiency of $50\%$ the  $q\bar{q} (\gamma)$ mistag rate 
is $0.8\%$ using combined $b$-tagging, while it is reduced to $0.5\%$ 
using equalised $b$-tagging. The $WW$ efficiency is
reduced from $0.1\%$ to $0.06\%$ 
when changing from combined to equalised tagging. 
The $ZZ$ efficiency is also reduced 
from $3.7\%$ to $3.3\%$. 

In the search for the SM Higgs boson
in the four jet channel, the $b$-tagging variable was 
combined with another set of discriminant variables\cite{fastpaper}
using an artificial neural network.
The final confidence level estimation is calculated 
using two-dimensional information,
where one dimension is the neural network output  and the other 
is the reconstructed Higgs boson mass. 

$b$-tagging is also used in the 
selection of the  Higgs di-jet \cite{fastpaper} from among the 6 possible
Higgs di-jet candidates in a 4-jet event.
The proportion of correct matchings for the Higgs di-jet,
estimated in simulated signal events with 114~GeV/c$^2$ mass,
is around 53\% at pre-selection level, increasing to above 70\% after the
tight selection cut (see table 1 of \cite{fastpaper}), while keeping a 
low rate of wrong pairings for $ZZ$~ background events.

\section{Conclusion}
\label{Conclusion}

The standard approach used by DELPHI for tagging $b$-hadrons has been 
described. By
using not only the track impact parameters, which are sensitive to the longer
lifetimes of hadrons containing $b$-quarks, but also other kinematic
information related to secondary vertices, track rapidities and any
leptons, the efficiency/purity has been improved. For $Z$ events, a purity of
98.5\% for $b$-jets was achieved for an efficiency of 30\%. Such high purity
was required for the accurate measurement of $R_b$ at the $Z$.

The tagging algorithm can also be applied to complete events, rather than
just to single hemispheres. For the SM Higgs search at LEP2, the sum of the
two largest $b$-tag variables for jets in the event was used. 
High efficiency $b$-tagging was required in order to extract any possible $H \to
b\bar{b}$ decays from the large backgrounds. For a signal efficiency of 60\%, a
rejection factor of 140 for the $W^+ W^-$ background was achieved.

In contrast, the algorithm could be used in an anti-tagging mode, to select
jets from light-quarks. This was used (together with conventional 
$b$-tagging), to compare the 3-jet rates for $b$- and for light 
quarks. This is sensitive to the $b$-quark mass.

For these and for other physics processes, it was crucial to have an efficient,
well understood procedure for tagging $b$ quarks. This resulted in systematic 
errors being kept to a minimum, and enabled many significant measurements to be
performed.

\subsection*{Acknowledgements}
\vskip 3 mm
 We are greatly indebted to our technical 
collaborators, to the members of the CERN-SL Division for the excellent 
performance of the LEP collider, and to the funding agencies for their

support in building and operating the DELPHI detector.\\
We acknowledge in particular the support of \\
Austrian Federal Ministry of Education, Science and Culture,
GZ 616.364/2-III/2a/98, \\
FNRS--FWO, Flanders Institute to encourage scientific and technological 
research in the industry (IWT), Belgium,  \\
FINEP, CNPq, CAPES, FUJB and FAPERJ, Brazil, \\
Czech Ministry of Industry and Trade, GA CR 202/99/1362,\\
Commission of the European Communities (DG XII), \\
Direction des Sciences de la Mati$\grave{\mbox{\rm e}}$re, CEA, France, \\
Bundesministerium f$\ddot{\mbox{\rm u}}$r Bildung, Wissenschaft, Forschung 
und Technologie, Germany,\\
General Secretariat for Research and Technology, Greece, \\
National Science Foundation (NWO) and Foundation for Research on Matter (FOM),
The Netherlands, \\
Norwegian Research Council,  \\
State Committee for Scientific Research, Poland, SPUB-M/CERN/PO3/DZ296/2000,
SPUB-M/CERN/PO3/DZ297/2000, 2P03B 104 19 and 2P03B 69 23(2002-2004)\\
JNICT--Junta Nacional de Investiga\c{c}\~{a}o Cient\'{\i}fica 
e Tecnol$\acute{\mbox{\rm o}}$gica, Portugal, \\
Vedecka grantova agentura MS SR, Slovakia, Nr. 95/5195/134, \\
Ministry of Science and Technology of the Republic of Slovenia, \\
CICYT, Spain, AEN99-0950 and AEN99-0761,  \\
The Swedish Natural Science Research Council,      \\
Particle Physics and Astronomy Research Council, UK, \\
Department of Energy, USA, DE-FG02-01ER41155. \\


\newpage

\clearpage

\end{document}